\documentclass{emulateapj}
\usepackage{lscape}

\newcommand{\kms}{\,km~s$^{-1}$}     \newcommand{\sqcm}{\,cm$^{-2}$}  
\newcommand{\fuse}{\emph{FUSE}}      \newcommand{\hst}{\emph{HST}}
\newcommand{\os}{\ion{O}{6}}         \newcommand{\ct}{\ion{C}{3}}
\newcommand{\hi}{\ion{H}{1}}
\newcommand{\hv}{high-velocity}
\newcommand{\Hv}{High-velocity}
\newcommand{\hva}{high-velocity absorption}
\newcommand{\hvr}{high-velocity absorber}
\newcommand{\hvo}{high-velocity \ion{O}{6}}
\newcommand{\hvh}{high-velocity \ion{H}{1}}
\newcommand{\hpv}{high-positive-velocity}
\newcommand{\pvw}{positive-velocity wing}
\newcommand{\Pvw}{Positive-velocity wing}
\newcommand{\pvc}{positive-velocity component}
\newcommand{\nvc}{negative-velocity component}

\begin{document}
\shorttitle{\os, \ct, and \hi\ in HVCs}
\shortauthors{Fox et al.}
\title{A Survey of \os, \ct, and \hi\ in Highly Ionized
  High-Velocity Clouds} 
\author{Andrew J. Fox\altaffilmark{1}, Blair D. Savage, \& Bart P. Wakker}
\affil{Department of Astronomy, University of Wisconsin -
Madison, 475 North Charter St., Madison, WI 53706}
\altaffiltext{1}{Current address: Institut d'Astrophysique de Paris, 98
  bis, Boulevard Arago, 75014, Paris, France}

\begin{abstract}
 We present a {\it Far-Ultraviolet Spectroscopic Explorer}
 survey of highly ionized high-velocity clouds (HVCs) in 66
 extragalactic sight lines with S/N$_{1030}>$8. 
 We searched the spectra for
 \hv\ ($100<|v_{LSR}|<400$\kms) \os\ absorption
 and found a total of 63 absorbers, 16 with 21\,cm-emitting \hi\
 counterparts and 47 ``highly ionized'' absorbers without 21\,cm emission.
 The highly ionized HVC population is characterized by
 $\langle b$(\os)$\rangle$=38$\pm$10\kms\ and
 $\langle$log\,$N_a$(\os)$\rangle$=13.83$\pm$0.36, with
 negative-velocity clouds generally found at $l<180$\degr\ and
 positive-velocity clouds found at $l>180$\degr. 11 of these \hvo\
 absorbers are 
 \pvw s (broad \os\ features extending asymmetrically to velocities of
 up to 300\kms).
 We find that $81$\% (30/37) of \hvo\ absorbers have
 clear accompanying \ct\ absorption, and $76$\% (29/38) have
 accompanying \hi\ absorption in the Lyman series.
 We present the first (\os-selected) sample of \ct\ and \hi\
 absorption line HVCs, and find 
 $\langle b$(\ct)$\rangle$=30$\pm$8\kms, log\,$N_a$(\ct) ranges from
 12.8 to $>$14.4, 
 $\langle b$(\hi)$\rangle$=22$\pm$5\kms, and log\,$N_a$(\hi) ranges from
 15.1 to $>$16.9.
 The lower average width of the \hvh\ absorbers implies the
 \hi\ lines arise in a separate, lower temperature phase than the \os. 
 The ratio $N_a$(\ct)/$N_a$(\os) is generally constant with
 velocity in \hvo\ absorbers, suggesting that \ct\ resides in
 the same gas as the \os. 
 Collisional ionization equilibrium models 
 with solar abundances can explain the \os/\ct\
 ratios for temperatures near 1.7$\times10^5$\,K; non-equilibrium
 models with the \os\ ``frozen-in'' at lower
 temperatures are also possible. Photoionization models are not viable
 since they under-predict \os\ by several orders of magnitude.
 The presence of associated \ct\ and \hi\ strongly
 suggests the \hvo\ absorbers are not formed in the hotter plasma that
 gives rise to \ion{O}{7} and \ion{O}{8} X-ray absorption.
 
 We find that the shape of the \os\ \pvw\ profiles is well reproduced by a
 radiatively cooling, vertical outflow moving with ballistic dynamics,
 with $T_0=10^6$\,K, $n_0\approx2\times10^{-3}$\,cm$^{-3}$,
 and $v_0\approx250$\kms.
 However, the outflow has to be patchy and out of equilibrium to
 explain the sky distribution and the simultaneous presence of \os,
 \ct, and \hi. 
 We found that a spherical outflow can produce \hvo\
 {\it components} (as opposed to the wings), showing that the
 possible range of outflow model results is too 
 broad to conclusively identify whether or not an outflow has left its
 signature in the data. 
 An alternative model, supported by the similar multi-phase structure
 and similar \os\ properties of highly ionized and 21\,cm HVCs, is one where
 the highly ionized HVCs 
 represent the low $N$(\hi) tail of the HVC population, with the \os\ formed at
 the interfaces around the embedded \hi\ cores. Though we cannot rule out the
 possibility that some \hvo\ absorbers exist in the Local Group or
 beyond, we favor a Galactic origin. This is based on the
 recent evidence that both \hi\ HVCs and the million-degree gas
 detected in X-ray absorption are Galactic phenomena. 
 Since the highly ionized HVCs
 appear to trace the interface between these two Galactic phases, it follows
 that highly ionized HVCs are Galactic themselves.
 However, the non-detection of \hvo\ in halo star spectra implies
 that any Galactic \hvo\ exists at $z$-distances beyond a few kpc.
\end{abstract}

\section{Introduction}
The {\it Far-Ultraviolet Spectroscopic Explorer} (\fuse) survey of
\os\ in the Galactic halo and high-velocity clouds \citep[HVCs;][ hereafter 
S03]{Wa03, Sa03, Se03} found that high-velocity  ($|v_{LSR}|>100$\kms)
\os\ is detected with a strength 
$W_{\lambda}>30$\,m\AA\ in 70\% of sight lines across the entire sky.
These absorbers trace gas created by a variety of processes.
The \hvo\ associated with known high-velocity \hi\ structures
(Complexes A and 
C, and the Magellanic Stream) has been well studied
\citep[S03,][]{Fo04,Fo05}. In these cases, the \os\ observations are best
explained by conductive or turbulent interfaces, where hot gas exists
in transition-temperature boundary layers between the cool/warm \hi\
HVCs and a surrounding hot ($\sim$10$^6$\,K) medium. In other
\hvo\ absorbers, no \hi\ 21\,cm emission is detected at the same
velocity as the \os; these absorbers are referred to as highly ionized HVCs
\citep{Se95,Se99,Co04,Co05,Ga05}. 
Ultraviolet (UV) and far-UV observations are required to detect and
study the material in highly ionized HVCs. 
A particular category of highly ionized HVC first described in S03 is the 
\pvw\  -- a broad feature extending asymmetrically 
to velocities of up to +300\kms\ without corresponding absorption at
negative velocity. 
Wings appear to be a continuation of low-velocity Milky Way absorption,
rather than discrete absorption components. Their nature and origin
is currently unknown. 

One interpretation of the \os\ wings (and potentially other \hvo\
absorbers) is that they trace a hot Galactic
wind or fountain. In this scenario, activity in the Galactic disk
(Type II supernovae, massive star stellar winds, and
possibly AGN activity) creates large volumes of hot, pressurized
interstellar gas that rises upward into the halo.
At low outflow speeds, the material will fall back to
the plane as a fountain \citep{SF76, Br80, HB90}.
A faster outflow (a wind) can escape the
Galactic potential well, in doing so 
transporting mass, metals, and energy to the
intergalactic medium \citep[see review by][]{Ve05}.
In our own galaxy,
mid-infrared images of the Galactic center show a large-scale bipolar
outflow \citep{BC03}. A lack of neutral hydrogen above $z\approx1$\,kpc in
the inner 3\,kpc of the Milky Way \citep{Lo84} supports the idea of a
wind, since the \hi\ could have been expelled or ionized. 
\os-containing outflows
are predicted by hydrodynamical simulations of galaxy winds
\citep{MF99, SS00}, and have been detected in starburst galaxies
NGC~1705 \citep{He01} and NGC~625 \citep{Ca04}. 
Although these these extragalactic absorbers exist in
the form of discrete components, not wings, they may appear to resemble wings
when observed from inside the galaxies. 
However, a hot Milky Way outflow has 
never been conclusively detected in the ultraviolet, but if present
should be detectable in \os\ absorption in the spectra of distant quasars.

Alternatively, many \hvo\ absorbers may arise at
large distances, rather than being associated with the Milky Way. The
possibility of large quantities of hot plasma 
pervading the Local Group has gained attention due to 
several detections of zero redshift absorption in the
X-ray lines of \ion{O}{7} and \ion{O}{8}
\citep[e.g.][]{Ni02,Fa03,Mc04,Wi05,Wi06}. This  
hot gas is likely tracing either the Warm-Hot Intergalactic Medium
(WHIM), or an 
extended Galactic corona (S03). In the WHIM interpretation, the X-ray
observations represent the detection of a warm-hot phase predicted by 
large-scale cosmological simulations \citep{CO99,Da01}. Such a phase
would contain a considerable fraction of the baryons at the current epoch
\citep[see][]{Ni02,Nc05}. However, the hot
extended corona interpretation is favored by  
observations of interstellar \ion{O}{7} absorption toward X-ray
binaries in a Galactic globular cluster \citep{Fu04} and in the LMC
\citep{Wa05}.

In this paper we extend the S03 survey by focusing on the highly
ionized HVCs, and looking for absorption in \ct\ and \hi\ accompanying
the \hvo. We use a new decontamination technique for removing the
effects of interstellar H$_2$ \citep{Wa06}, and
include new data taken with the \fuse\ satellite up to December 2004. The 
scientific motivation for this extension is to understand the origin
of the \pvw s, and the highly ionized HVCs in general. 

Section 2 describes our observations and data reduction. In \S3 we
describe the measurement of \hva, present column densities for
\os, \ct, and \hi, and display the spectra. In \S4 we analyze our
results, looking at kinematics and the distribution of \hvo\ on the 
sky, distance information, ionization, and the relationship between
absorption in different species. Section 5 contains 
a comparison of the properties of \hvr s with the predictions of
origin models. Our work is summarized in \S6.

\section{Observations and Data Handling}
\subsection{Data Reduction}
Our far-UV spectra were obtained between the years 2000 and 2005 using the
\fuse\ satellite 
\citep{Mo00, Sa00}. For wavelengths below 1000\,\AA\ (\ct\ and \hi\
lines), we use data from the SiC2A detector segment. For wavelengths
between 1000 and 1100\,\AA\ (\os\ lines), we use combined LiF1a and
LiF2b data. 

We extracted from the \fuse\ archives spectra of all extragalactic
sight lines 
publically available in May 2005 with signal-to-noise per
resolution element at 1030\,\AA\
(S/N$_{1030}$) $>$8\footnotemark. This produced a
list of 88 sight lines. We did not include
absorbers that may be present in data with S/N$_{1030}<$8, since
apparent column density measurements in data that noisy tend to be
overestimated \citep{FSW05}, and the H$_2$ absorption is difficult to
model \citep{Wa06}.
\footnotetext{The S/N is measured by calculating the rms dispersion
  around the fitted continuum using data rebinned to pixels
  equal in size to the resolution element (20\kms).}
The raw sample contains 55 directions at 
$b>0$\degr, and 33 directions at $b<0$\degr. A general absence of
sources in the range $|b|<20$\degr\ is due to 
high extinction in these directions. 

For each sight line the data were reduced with the calibration pipeline 
CALFUSE v2.1 or v2.4, with a further velocity zero-point
shift for each segment and for each exposure determined and applied as
described in \citet{Wa06}. 
Since only marginal
improvement in the calibration results from using v2.4 rather than v2.1, the
re-calibration of all datasets with a common pipeline version was
deemed unnecessary.

Continua were fit near the
lines of interest (\os\ $\lambda$1031.926, \ct\ $\lambda$977.020, and
the \hi\ Lyman lines) using Legendre 
polynomials. In cases where \ion{O}{1}$^*$ 
airglow emission is strong near 0\kms\ in \ct\ 
$\lambda$977.020, we use the night-only data for this line. Night-only
\fuse\ data is always used when 
measuring the \hi\ absorption in the $\lambda$972.537, $\lambda$949.743,
$\lambda$937.804, and $\lambda$930.748 lines,
unless the S/N in the night-only data is so poor that no useful
measurement is possible.

\subsection{H$_2$ Decontamination}
We applied a H$_2$ modelling
procedure to the \fuse\ data, using all available lines
from the Lyman and Werner bands \citep[for a full description of this
  process, see][]{Wa06}. The process characterizes each
component of H$_2$ absorption with six parameters: central velocity, line
width, total H$_2$ column density, and three excitation temperatures
that together give the relative populations for rotational 
levels $J$=0 to 4. This model  
is used to produce a new continuum with the effects of H$_2$ removed.,
In cases where multiple components of H$_2$ absorption exist, 
each component is fitted separately.

Of particular note in this study are the H$_2$ (6--0) P(3) and (6--0)
R(4) lines at 1031.191 and 1032.356\,\AA, respectively.
These features lie at $-$214 and +123\kms\ in the frame of \os\
$\lambda$1031.926, and so could mimic the 
effect of \hvo\ absorption. The precision of our $N$(\os) measurements
is thus limited by the quality of the H$_2$ modelling, with the error
on $N$(\os) including a systematic contribution from the uncertainty on
the H$_2$ model.

\subsection{Removal of Contaminated Spectra}
We omitted 22 sight lines where we have no information on the presence
of HVCs in absorption in \os\ $\lambda$1031.926.   
The reasons for this were:
blends with intrinsic or intergalactic absorption
line systems, highly uncertain continuum placement, or complicated H$_2$
lines that could not be reliably modelled.
Table 1 contains a list of the 66 remaining sight lines, together with a
summary of their basic information. 

\subsection{Identification of Highly Ionized HVCs and \hi\ HVCs}
We searched the spectra in the sample
for \hva\ in \os\ $\lambda$1031.926, defined in the 
range $100<|v_{LSR}|<400$\kms, and found 49/66 sight lines showed \hvo\
detections. \os\ absorption in the range $-$100 to 
100\kms\ is usually attributed to the Galactic ``thick disk'' \citep{Sa03},
whereas \os\ absorption in the range $400<|v_{LSR}|<1000$\kms\ is
rare, and appears to be associated with external galaxies (S03). In
most cases \os\ $\lambda$1037.617 was not useful for studying
\hvo\ due to blends with \ion{C}{2} $\lambda$1036.337, \ion{C}{2}$^*$
$\lambda$1037.018, H$_2$ $\lambda$1037.149, and
H$_2$ $\lambda$1038.157.

Since the \os\ associated with 21\,cm \hi-emitting HVCs has been well
studied, we then excluded from our survey 16 \hvo\ absorbers clearly
associated (both spatially and kinematically) with 
Complex A, Complex C, the Outer Arm, or the Magellanic 
Stream. For the purpose of this exercise, the $2\times10^{18}$\sqcm\ \hi\
21\,cm contour \citep{HW88} was used to define the edge of the HVCs,
and any \hvo\ absorbers seen in sight lines passing through these
clouds at the velocity of the \hi\ HVC were judged to be associated
with the \hi\ HVC. 
The sight lines with excluded absorbers are not discarded, since they
contain valuable information on \os\ absorption (or lack thereof) at
other velocities. 

The sky distribution of our sample can be seen in Figure 1, showing
the \hvo\ sky at both negative and positive velocities. The 
circle size is scaled by the \os\ column density in the high-velocity
absorber (see \S3 for a description of how $N$(\os) is measured).

\subsection{Incidence of \Hv\ \os\ Absorption}

\begin{itemize}
\item \Hv\ \os\ absorption appears in 74\% (49/66) of extragalactic
  \fuse\ spectra with S/N $>8$. This compares to the findings of S03
  who detected \hvo\ in 84\% (84/100) of extragalactic sight lines
  (including lower S/N data).
\item 23\% (15/66) of sight lines show \hvo\ with 21\,cm \hi\ counterparts.
  These are the ``regular'' HVCs with log\,$N$(\hi)$\gtrsim18.2$. 
\item 52\% (34/66) of sight lines show \hvo\ with no 21\,cm \hi\
  counterparts. These are the highly ionized HVCs. 
\item 26\% (17/66) of sight lines show no detections of \hvo.
\item In the 15 sight lines showing \os\ absorption associated with
  21\,cm emitting clouds, 16 \hvo\ components are found.
  ESO265-G23 is the single case where there appear to be two \hvo\
  components, in the velocity range of the Magellanic Stream.
\item In the 34 sight lines showing highly ionized \os, a total of 47 
  highly ionized HVCs are found (20 North and 27 South,
  30 at positive velocity and 17 at negative velocity; 22 sight lines
  show one
  \hvo\ absorber, 11 sight lines show two, and one sight line shows
  three.
\item Approximately three out of four \hvo\ absorbers are highly
  ionized, i.e. do not have 21\,cm counterparts.
\item Since there could be highly ionized \os\ masking behind \os\
  associated with 21\,cm HVCs, the number of uncontaminated sight
  lines in which to look for highly ionized HVCs is 66--15=51.
  Thus, the corrected incidence of highly ionized HVCs is 
  47 in 51 sight lines with
  no \hv\ \hi\ 21\,cm emission (0.92 highly ionized HVCs per sight line).
\end{itemize}

Our sample of highly ionized HVCs is biased by the
exclusion of absorbers centered in the range $|v_{LSR}|<100$\kms, and
by the removal of velocity ranges in directions showing strong HVC 21\,cm
emission. 
Many of the sight lines were also originally selected to be observed
based on a 
low $N$(\hi), for low extinction; this produces a separate
observational bias.

\section{Measurement of Absorption}
We used the apparent optical depth (AOD) method \citep{SS91,SS92} to measure
the apparent column densities of \os, \ct, and \hi\ absorption in the
\os\ HVCs. 
This technique requires no prior knowledge about the component
structure and is valid for measurement of velocity-resolved,
unsaturated lines.
The high-temperature environments in which \os\ is thought to exist
ensure that the \os\ lines have enough thermal broadening to be
resolved. Saturation can be checked for using the weaker \os\ line,
but numerous blends at high-velocity prevent the assessment being
made in our dataset. \citet{Sa03} demonstrated that saturation was
generally not a 
problem for the thick-disk \os\ absorption in a similar-sized
\fuse\ dataset that included many of the same sight lines we study
here; by association, the \hvo\ in our survey, which is generally
weaker than the thick-disk \os, is likely unsaturated too. 

The main spectroscopic measurements for sight lines with \hvo\
detections are presented in Table 2. 
For each sight line we list the range of high-velocity absorption
(identified by eye using the \os\ line),
then for \os, \ct, and \hi\ we list the central velocity $v_0$,
line width $b$, equivalent width $W_{\lambda}$ in the velocity range
$v_-$ to $v_+$, apparent column density $N_a$ between $v_-$ and
$v_+$, and detection significance, defined as
$W_{\lambda}/\sigma(W_{\lambda})$. For \hi, the Lyman series
absorption line chosen to make the 
measurement is listed. This choice is made by finding a relatively
weak line free from airglow and blends. 
For cases where \hvo\ is seen with no
counterpart in \ct\ or in \hi, we present 3$\sigma$ upper limits to
log\,$N$ in these species, using the 3$\sigma$ equivalent
width limit in the same velocity integration range used for
\os, and a linear curve-of-growth. In Table 3 we list those sight
lines and velocity ranges without \hvo\ detections and the
corresponding 3$\sigma$ upper limits for $N$(\os). 

Measuring $N$(\hi)
in the \hvr s is much more challenging than measuring \os, for the
following reasons: 
(1) Noisy data in the SiC channels: we find S/N$_{970}$ is typically
0.5S/N$_{1030}$. 
(2) Saturation: many \hvr s are saturated through the weakest
Lyman line observable with \fuse\ (\hi\ $\lambda$917.181).
(3) Blending: contamination arises from a high concentration of H$_2$,
\ion{O}{1}, and other lines.
(4) Atmospheric emission: airglow is seen near 0\kms\ in the stronger Lyman
lines. The airglow, which can be
reduced but not fully removed by using night-only rather than combined
day+night data, leads to a shortage of information
on the shape of the interstellar line profile in the range
$|v_{LSR}|\lesssim100$\kms.
(5) Difficulties in continuum placement: the high density of
absorption lines at $\lambda<1000$\,\AA\ results in a lack of spectral
regions free from absorption, needed to place the continuum.
(6) Assuming the \hi\ lines are resolved: the AOD technique can only
be applied to lines that are resolved, otherwise the column densities
will be underestimated. If the \hi\ exists in cool gas at
$T<1.4\times10^4$\,K and has no non-thermal 
broadening, the Lyman lines will not be resolved by the \fuse/SiC
channels, and then full saturation could be hidden. 
For these reasons, the uncertainty on $N$(\hi) is much larger than
on $N$(\os) or $N$(\ct). However, even without precise measurements of
$N$(\hi), we can still make an assessment of whether \hi\ absorption is
present at the velocity of the \os\ absorption.

Plots showing the \os, \ct, and \hi\ absorption for all the
highly ionized HVCs are presented in Figure 2.
We classified each absorber into one of three categories based on
the properties of the \hvo\ absorber: \pvw s (broad absorption
blending with the thick disk at $|v_{LSR}|<$100\kms\ and continuing
out up to 300\kms), \pvc s (isolated
absorption components), and \nvc s. No negative-velocity wings were
observed. Summary plots showing \os, \ct, and \hi\
for all absorbers in each category are displayed in
Figures 3 to 5. In Figure 6 we include all \hvo\ non-detections,
together with the profiles of those \os\ absorbers excluded from our
survey (those associated with 21\,cm \hi-emitting HVCs).

\section{Analysis}
\subsection{Distribution of Highly Ionized HVCs on the Sky}
Table 4 contains analysis of the properties and sky
distribution of highly ionized \hvo\ absorbers. In Figure 7 we show
the distribution of absorber velocities
with longitude and latitude, in both the Local Standard of Rest (LSR) and
Galactic Standard of Rest (GSR) reference frames
 ($v_{GSR}=v_{LSR}+(220\,\mathrm{km\,s}^{-1})\,{\mathrm{sin}}\,l\,{\mathrm{cos}}\,b$.) 
The overall distribution has a marked dipole appearance, as evidenced
by the following:
\begin{itemize}
\item 82\% of negative-velocity highly ionized HVCs are at $l<180$\degr.
\item 94\% of negative-velocity highly ionized HVCs are at $b<0$\degr.
\item 73\% of positive-velocity highly ionized HVCs absorbers are at
  $l>180$\degr. 
\item 63\% of positive-velocity highly ionized HVCs absorbers are at
  $b>0$\degr. 
\end{itemize}

In the following points we discuss several features of
the distribution of \hvo\ on the sky (Figs. 1 and 7).

{\it There is a lack of highly ionized HVCs, both at positive and
  negative velocities, in the quadrant $l<180$\degr, $b>0$\degr.} Both
  selection effects and real effects appear to be at work here; these
  are discussed in turn in the next two points.

{\it Two selection effects prevent the detection of highly ionized HVCs in
  the range} 0\degr\ $<l<180$\degr, 0\degr\ $<b\lesssim45$\degr.
  At {\it negative} velocities the lack of detections is largely due to
  the exclusion of sight lines through Complex~C and Complex~A. 
  There could be negative-velocity highly ionized HVCs in these
  directions, but we have no way of separating these absorbers from
  \os\ absorption associated with the 21\,cm-emitting structures.
  At {\it positive} velocities, the lack of highly ionized HVCs in
  this region may be related to the LSR motion around the Galactic
  center. For example, material moving at 
  $v_{GSR}$=100\kms\ along sight lines at 0\degr\ $<l<180$\degr\ and
  low Galactic latitude would
  be masked, from our vantage point, by thick disk 
  material since the solar motion is in this direction; 
  our absorber selection procedures would therefore not identify
  such material as high-velocity. 
  This argument can also explain the non-detections of highly ionized
  HVCs in the region 0\degr\ $<l<180$\degr, $-$45\degr\ $\lesssim b<0$\degr.

{\it There are several significant non-detections of \hpv\ \os\ at} 
  0\degr\ $<l<180$\degr, $b\gtrsim$ 45\degr\ (e.g. Mrk~279, PG~1259+593). 
  At higher latitude,
  the component of solar motion decreases, so any outflowing material
  in the first two quadrants will no longer be masked by 
  rotating disk material in the foreground. The absence of
  high-latitude \hpv\ \os\ absorbers at 0\degr\ $<l<180$\degr\ (the
  Complex C sight lines) is thus an important observation,
  particularly because the \fuse\ data in these sight lines tend to
  have good S/N ratios. 
  Mrk~817 is the only sight line showing \hpv\ \os\ in this part of
  the sky\footnotemark. 
  \footnotetext{Though the Mrk~817 features (two components are seen) are
  weak, the spectrum has very high 
  signal-to-noise and the 190\kms\ component is also seen in
  \os\ $\lambda$1037.617.}


{\it Many \hvo\ absorbers are close to the orbit of the Magellanic
  Stream.} We note the similarity in kinematics and location on the
  sky between many \hvo\ absorbers and the Magellanic Stream (see Figure
  1). If many \hvo\ absorbers were associated with the Stream, it
  would indicate the Stream extends further out in space than the
  regions sampled by \hi\ 21\,cm emission \citep[see][ for further
  evidence of this conclusion]{Fo05}. Note that the \nvc s seen in the range
  $20$\degr\ $<l<120$\degr, $-60$\degr\ $<b<-30$\degr\ and
  $v\approx-150$ to $-$100\kms\ 
  appear {\it unrelated} to the Magellanic Stream despite their proximity on
  the sky, since the Stream has velocities over 100\kms\ more negative
  than the absorber velocities in these
  directions. 

{\it In sight lines close to but off the edge of 21\,cm complexes, the
  highly ionized boundary layers appear to be fairly confined.} 
  Note the 
  high-negative-velocity \os\ non-detections in sight lines close to but
  a few degrees outside the \hi\ boundaries of Complex C and the
  anti-center clouds (Figure 1). The \os\ associated with 
  these 21\,cm-emitting complexes is therefore confined rather than
  extended, supporting the idea that the large complexes are surrounded
  by relatively narrow \os-bearing interface layers \citep{Fo04}. In the
  conductive interface models of \citet{Bo90}, the typical thickness
  of an \os-bearing conduction front is 15\,pc (or 5\arcmin\ for a
  cloud at 10\,kpc).

\subsection{Presence of Other Ions}
Among the more important results of our survey 
is that approximately 80\% of \hvo\ absorbers have clear counterparts
in both \ct\ and the \hi\ Lyman series.\footnotemark
\footnotetext{When determining the  
fraction of \os\ absorbers with clear \ct\ or \hi\ counterparts, we
exclude cases where blends or low-quality data prevents the assessment
from being made; these cases thus do not contribute to the statistics. }
 In Figure 8 we display the sky distribution of HVC absorption in \os,
\ct, and \hi, showing the directions in which all three ions are detected
over the same velocity range. The seven cases where \hv\ \ct\ is not
detected at 3$\sigma$ levels are 
Mrk~817 (189\kms\ component), Mrk~421, PG~1302--102, PG~1011--040
(279\kms\ component), MS~0700.7+6338, 
HE~1143-1810 (246\kms\ component), and Ton~S180 (251\kms\
component). 
The nine cases where \hvh\ is not detected, namely 3C~273.0, Mrk~817
(189\kms\ component),  
Mrk~421, Mrk~1383, ESO141--G55, PG~1302--102, MS~0700.7+6338,
HE~1143-1810 (246\kms\ component), and Ton~S180 (251\kms\ component) are
sight lines with low column density \hvo: eight of these nine
absorbers have log\,$N$(\os) lower than the overall average
log\,$N$(\os) of 13.87. In other words, all strong \hvo\ absorbers
have \ct\ and \hi\ counterparts.

Though these are \os-selected samples, they represent the first
systematic study of \ct\ and \hi\ absorption in HVCs.
Among the population of 14 unsaturated \ct\ absorbers, we find
$\langle b$(\ct)$\rangle$=30$\pm$8\kms; log\,$N_a$(\ct) ranges from
12.8 to $>$14.4.
Among the population of \hva\ apparently unsaturated in \hi\ (11
absorbers), we find $\langle b$(\hi)$\rangle$=22$\pm5$\kms;
log\,$N$(\hi) lies in the range 15.1 to $>$16.9. 
We show in Figure 9 histograms of $v_0$, $b$,
and log\,$N$ for the \os, \ct, and \hi\ absorbers, by absorber
category (wing or component). 
The \hi\ absorption line HVCs are significantly narrower, and the \ct\
HVCs somewhat narrower, than their \os\ counterparts. 

Figure 10 contains scatter plots showing a comparison of the \os\
properties of HVCs with the corresponding \ct\ and \hi\ properties.
The top panels clearly show a strong kinematic correspondence between
all three ions. 
The middle and lower panels, comparing line widths and column
densities, show an interesting property: there
are trends for $b$(\ct) to trace $b$(\os) (linear correlation coefficient =
0.70) and log\,$N$(\ct) to trace log\,$N$(\os) (linear correlation
coefficient = 0.71), but log\,$N$(\hi) and log\,$N$(\os) are
uncorrelated (linear correlation coefficient = 0.05). 

\subsection{Wings vs Components}
23\% of \hvo\ absorbers (and 37\% of high-positive-velocity \os\
absorbers) exist in the form of \pvw s. The wings are
always seen at positive velocity, 8/11 at $b>0$\degr, and all but two
are found at $l>179$\degr\ (exceptions: Mrk~509 and Mrk~817). 
Wings and components can each be weak or strong, with log\,$N$(\os)
between 13.0 and 14.5 in both cases. 

In Table 5 we look for differences in the average properties of \pvw
s, \pvc s, and \nvc s. The calculation of ionization
properties, using the ionic ratios \ct/\os\ and \hi/\os, ignores cases
where \ct\ or \hi\ is not detected.
The \pvc s are slightly narrower and have lower average \ct/\os\
and \hi/\os\ ratios than \pvw s. 
There is also a higher incidence of non-detections of \ct\ and
\hi\ among \pvc s: wings
show accompanying \ct\ and \hi\ at 3$\sigma$ levels in 8/9 cases; \pvc s
show \ct\ in 13/19 cases and \hi\ in 11/19 cases. 
However, apart from the likelihood of a \ct\ or \hi\ counterpart, none
of the wing/component differences are statistically
significant. Furthermore, if one considers the \nvc s 
and \pvc s together, then their
distributions of \os\ column density, width, 
and ionization properties become similar to those of wings. 

We conclude that there is no strong evidence for a physical difference
between wings and components.
The similarity of the wing population to the
component population can be seen in the histograms shown in Figure 9.

\subsection{Comparison with Earlier Results}
Our sample of 11 \os\ \pvw s is significantly different from the sample of
22 \pvw s in S03. We find that the reported \os\ features toward
Mrk~876 and 3C~273.0 (125\kms\ component) are likely due to H$_2$
contamination, and the reported 
wings toward PG~1351+640 and PG~1259+593 are not significant at the
3$\sigma$ level. We have
reclassified the absorbers toward NGC~1705, 3C~273.0 (210\kms\
component), and Mrk~1383 as components, not wings. The S03
wings toward PG~0947+396, HS~1102+3441, PG~1001+291, Mrk~734,
HE~1115--1735 and Tol~1247--232 have too low S/N to be included in our
data set. Finally, we have included the wings toward PG~1011-040,
NGC~625, and NGC~5408 in data that were released after the publication
of the S03 survey.

Our non-detection of \hv\ \ct\ absorption toward Mrk~421 is
in contrast to \citet{Sa05}, who claim a detection of
$W_{\lambda}=26\pm10$\,m\AA\ at 60--165\kms. This absorber is not
included in our sample because we only consider 3$\sigma$ detections.
 
\subsection{Constraints on the Distance to Absorbing Gas}
If the \os\ wings (and highly ionized HVCs in general) do trace
outflowing Galactic material,
they should be visible in the spectra of Galactic halo stars at high
$z$-distance. \citet{Zs03} found no evidence for \os\ wings in 22
sight lines passing through the Galactic halo, with the exception of
HD 100340, where substantial continuum placement uncertainty allows
for the possible presence of \hvo. \citet{Zs03} point
out that if \hvo\ were as common towards halo stars as it is toward
extragalactic targets, it would appear in 10--12 sight
lines in their sample. However, we note that 
if the gas that gives rise to broad, shallow absorption
wings were present toward halo stars, it would be difficult to detect
against the complex, undulating, hot stellar continua. 

The sight lines to BD+38~2182 and Mrk~421 form an
interesting case of a halo star (BD+38~2182) closely aligned with an
extragalactic target (Mrk~421), both having been
observed with \fuse\ \citep{Sa05}. BD+38~2182, a B3V star at $z=3.5$\,kpc
does not show an \os\ \pvw, even though it lies
$\approx4$\degr\ away from the sight line to AGN Mrk~421, whose spectrum
displays an \os\ wing extending to 160\kms. 
In another halo star/AGN pair, there is weak \os\ wing absorption in
the range 
80 to 160\kms\ in the spectrum of halo star HD~100340 (even allowing
for molecular hydrogen at 125\kms), but much
stronger absorption 
is seen in this velocity range along the nearby sight line to AGN Mrk~734.
Finally, vZ~1128, a globular cluster star at
$z\approx$10\,kpc, shows no \hvo\ absorption \citep{Ho03}.

All these indicators (shown in Figure 11), 
suggest the \hvo\ originates beyond a few kpc, though it is impossible
to draw conclusions on the whole sky from a small number of cases.
If the \hvo\ existed in an extended
distribution surrounding the galaxy, the fraction of \hvo\ absorbers in
the inner few kpc (and hence intercepted by the halo star sight
lines) could be small.
On the other hand, the lack of \hvo\ detections towards halo stars has
been used to argue for an extra-galactic origin for all the \hvo\
clouds \citep{Nc05}.

\subsection{Ionization of \hvo}
Considerable work on the ionization of interstellar \os, both in the
Galactic thick disk \citep{Sa03, Zs03} and in
high-velocity clouds \citep{Se03,Fo04,Fo05,Co04,Co05,Ga05} has
led to the understanding that \os\ cannot be photoionized in these
environments, because the
absorber sizes necessary to reproduce the observed HVC ionization pattern
given the Galactic and extragalactic radiation fields are far too
large to be reasonable, and inconsistent with the results
from modelling the ions \ion{C}{2}, \ion{C}{3}, \ion{Si}{2}, and
\ion{Si}{3}. Collisional processes are thus thought to 
dominate the production of \os\ \citep[see reviews in][]{Fo04, IS04}. 

Without \hst\ data for many of the objects in our sample, we do not have
\ion{C}{4} and \ion{Si}{4} measurements to help diagnose the ionization
mechanism for each individual case. However, if the absorbers are in
collisional ionization equilibrium (CIE), and a solar elemental
abundance pattern is assumed\footnotemark,
\footnotetext{We adopt
  $A_{\mathrm{O}}^{\odot}=(n_{\mathrm{O}}/n_{\mathrm{H}})_{\odot}=10^{-3.34}$ \citep{As04}, and 
  $A_{\mathrm{C}}^{\odot}=(n_{\mathrm{C}}/n_{\mathrm{H}})_{\odot}=10^{-3.61}$ \citep{AP02}.}
the integrated column density ratio $N$(\ct)/$N$(\os) 
can be used to 
estimate the implied temperature \citep{SD93}. For the \hvr s in this
study, these results are presented in Table 6. 
Typical \ct/\os\ ratios in wings imply log\,$T$ = 5.23, or
$T$=1.7$\times10^{5}$\,K, though since gas near these temperatures
cools very quickly, we consider thermal equilibrium to be unlikely.
Although CIE solutions also exist for the \hi/\os\ and \hi/\ct\ ratios,
the lower average width of the \hvh\ absorbers ($\langle
b$(\hi)$\rangle$=21$\pm$7\kms\ versus $\langle
b$(\os)$\rangle$=38$\pm$10\kms) implies that 
log\,$T$(\hi)$<$4.67, at which temperature no \os\ would form,  unless
the gas is in
a highly non-equilibrium state or is photoionized, which we have
already ruled out. {\it Thus, the \hi\ exists in a separate, lower
temperature phase than the \hvo.} This is also suggested by the lack
of correlations between $b$(\os) and $b$(\hi), and log\,$N$(\os) and
log\,$N$(\hi), shown in Figure 10\footnotemark.
\footnotetext{\citet{DS05} and \citet{DS06} found similar evidence
  for multiple gas phases in their studies of \os, \ct, and \hi\ in the
  low-redshift intergalactic medium.} 
The CIE temperatures derived from the \hi/\os\  (and also \ct/\hi) ratios
then becomes meaningless. 

More detailed information on ionization is provided by looking at
the apparent column density ratios as a function of velocity, rather
than the integrated values. This approach is 
problematic for ions with markedly different atomic weights and
thermal broadening, but valid for \os\ and \ct. We have derived
$N_a$(\ct)/$N_a$(\os) ratios as a function of velocity in our highly
ionized absorbers, for a selection of \pvw s, \pvc s, and \nvc s
(Figure 12). 
In each category of \hvo\ absorber, $N_a$(\ct)/$N_a$(\os) ranges
between 0 and 4. {\it In general, there is 
  no evidence for a slope in $N_a$(\ct)/$N_a$(\os) with velocity}, 
lending support to the idea that the \os\ and \ct\ ions reside
in the same gaseous phase. 
In a few cases (both wings and components) some peaks in the ratio
are seen, indicating relative enhancements of \ct. 
Such enhancements could be due to a photoionized \ct\ component
existing in addition to a collisionally-ionized component, or alternatively
from a lower-temperature \ct\ component associated with the \hi.

\section{Wing Origin Models}

\subsection{Galactic Outflows}
\os\ outflows are seen in galaxies NGC~1705 \citep{He01}
and NGC~625 \citep{Ca04}. 
The idea that the \os\ \pvw s may trace an analogous Milky Way outflow
has been suggested before \citep[][ S03]{Se01}. 
In this section we explore the outflow hypothesis by developing
one-dimensional models to determine the 
\os\ absorption line profile expected in a Galactic outflow.
We develop some useful insights into
the expected absorption line signature of a Galactic outflow without a
full hydrodynamical treatment \citep[for which we refer the reader
  to][]{HB90}. 

We consider outflows rather than inflows or rotating halos,
since these other models have been shown to have difficulties
explaining the \os\ observations. S03
showed that models of a co-rotating or static halo do not explain
the kinematics of the \hv\ \os\ absorbers, and infalling material
\citep{Co05} cannot explain the numerous positive-velocity absorbers in the
anti-center direction. 

We derive synthetic \os\ and \ct\
column density profiles produced when looking through a uniform,
radiatively 
cooling Galactic outflow, moving with
ballistic dynamics in the Galactic gravitational field. Two geometries
for the outflow are considered: vertical and spherical.
Our models are calculated numerically by following the evolution of
physical 
conditions on a grid with time interval of
1\,Myr. The free parameters (initial conditions) in the models are
the temperature, density, and velocity at the base of the flow 
($T_0$, $n_0$, and $v_0$).
To ensure a tractable problem, we restrict ourselves to
the directions $l=0$\degr\ and 180\degr, so that Galactic rotation does not
affect the line profiles\footnotemark\footnotemark.
\footnotetext{Fortunately, several observed directions (e.g. Mrk~421,
  PG~0953+414) lie very close to $l$=180\degr\ and have high S/N spectra.}
\footnotetext{See \citet{Co02} for an investigation on the effects of
Galactic rotation on the ballistic motion of upward-directed clouds.}

In the vertical case,
the equation of motion of a parcel of gas moving vertically upward is
given by 
$\frac{{\mathrm d}v(z)}{{\mathrm d}t}=-g(z)$,
where we take the gravitational acceleration perpendicular to the plane
from \citet[][Figure 1.7]{Sp78}, derived using the measured
$z$-distribution of spectral type K giants in the Galactic halo. 
We impose mass continuity by insisting that
$v(z)n(z)=v_0n_0=\mathrm{constant}$; this mass flux
can be converted to a total mass flux out of the Galaxy
by assuming uniform flow out of
a circular disk with radius 20\,kpc (i.e., assuming a covering
fraction of unity). 

In the radial case, we use a spherical Milky Way mass distribution 
truncated at $r_{max}$=50\,kpc \citep{BT87} to determine the
gravitational acceleration per unit mass:
$g(r) = -v_c^2/r$ for $r<r_{max}$ and $g(r) = -v_c^2(r_{max}/r^2)$ for
$r>r_{max}$. The condition of conservation of mass now becomes 
$r^2n(r)v(r)=r_0^2n_0v_0=\mathrm{constant}$.

Internal energy loss in the models is through radiative cooling:
$\frac{{\mathrm d}E(z)}{{\mathrm d}t}=-\Lambda(T[z])n_{\mathrm{H}}^2(z)$
where the cooling function $\Lambda(T)$ is taken from \citet{HB90} and
$n_{\mathrm{H}}$ is the total (H$^0$+H$^+$) hydrogen density in cm$^{-3}$.
Given the initial conditions, the equation of motion is solved to
yield $v(z)$, which combined with mass continuity gives 
$n(z)$, which can be used to derive $T(z)$ using the rate of
cooling. At each grid cell in our flow, we determine the fraction
of atoms in the cell that are O$^{5+}$ and C$^{2+}$
$n_{\mathrm{O~VI}}(z)=f_{\mathrm{O~VI}}[T(z)]A_{\mathrm{O}}n_{\mathrm{H}}(z)$
and
$n_{\mathrm{C~III}}(z)=f_{\mathrm{C~III}}[T(z)]A_{\mathrm{C}}n_{\mathrm{H}}(z)$,
where $f_{\mathrm{O~VI}}$ is the fraction of oxygen atoms existing as
\os\ and $A_{\mathrm{O}}$ is the oxygen abundance (with quantities for
carbon similarly defined). Here we have assumed
CIE ionization distributions at each temperature,
and solar oxygen and carbon abundances. 
The model is run until either the flow cools to 10$^4$\,K, at which
point no \os\ or \ct\ will be detectable, or the flow slows to zero,
where it will blend with gas in the thick disk. 
Our outflow models do not include expansion cooling. In a true 
three-dimensional flow, the gas could
expand laterally as it rises upward, doing work at the expense of
internal energy.
The effect of including expansion cooling, a secondary source of
energy loss, would be to increase the column density of \os\ at the
expense of \ion{O}{7}. We also do not consider the effects of
buoyancy, drag forces, or magnetic fields.

For a sight line at a given latitude, we interpolate the run of
physical quantities with $z$ (vertical model) or $r$ (spherical model)
onto distance along the sight line, then integrate along the sight
line to obtain the predicted column density profile. 
We add thermal broadening by
convolving the column density profile with a Gaussian having
$b=(2kT/m)^{1/2}$ (=32\kms\ for \os\ at 10$^6$\,K).
The column density profiles are converted to optical depth
profiles using the relation $\tau(v)=N(v)f\lambda/3.768\times10^{14}$,
where $\lambda$ is the transition wavelength in cm and $f$ the 
oscillator strength of the transition\footnotemark.
\footnotetext{\os\ $\lambda$1031.926, $f$=0.1325; \ct\
  $\lambda$977.020, $f$=0.7570 \citep{Mo03}.}
In order to make our simulated observation as realistic as possible,
we then generate a Gaussian thick disk \os\ component, characterized by
$v_0=0$\kms, 
$b=60$\kms, log\,$N$(\os)=14.34 \citep[][]{Sa03}, and a 
central optical depth $\tau_0=1.1497\times10^{-2}N\lambda f/b$
\citep{Sp78}. 
We add the disk $\tau(v)$ to the outflow $\tau(v)$ to produce the
total optical depth profile.
The properties of \ct\ absorption in the Galactic disk have not been fully
characterized, so we 
proceed by using a \ct\ disk component with identical $v_0$, $b$, and
log\,$N$ as the \os\ disk component.
To simulate the measurement of this gas with \fuse, we sample the profiles
onto 2.0\kms\ pixels, then generate a normalized flux profile using 
$F_{norm}(v)=e^{-\tau(v)}$. We convolve the flux profile with a
Gaussian line spread function to represent instrumental broadening by
the \fuse/LiF spectrograph (FWHM=20\kms). Finally, we add Poisson Noise
at a (typical) level of S/N per resolution element = 12. 

In Figure 13 we show the results from a typical model run. 
This model is parameterized by $T_0=10^6$\,K,
$n_0=2.0\times10^{-3}$\,cm$^{-3}$, and $v_0=250$\kms.
{\it The qualitative shape of the \os\ wings can be well explained by
the vertical Galactic outflow.} 
As the gas rises, it initially maintains a high temperature and level
of ionization since
the rate of radiative cooling is low. When the Galactic gravitational
field has slowed the flow to near zero velocity,
the density rises to conserve mass, and cooling starts to occur
rapidly, causing the temperature to drop to 10$^4$\,K.
These models therefore explain the observed behavior that wings are not
seen at negative $v_{LSR}$ (neglecting Galactic rotation effects).
The \os\ in these models traces million-degree gas rising into the
halo, not gas at temperatures near $3\times10^5$\,K where \os\ peaks
in abundance.

With the CIE ionization assumption, \ct\ wings should not be seen
since the flow is too hot to produce substantial quantities of \ct.
This prediction is clearly
at odds with our detection of \ct\ wings in 8/9 cases, so a CIE flow
is ruled out. 
A {\it smooth} Galactic outflow is
ruled out by the observations that wings are only seen in 14\%
of extragalactic sight lines, that a North/South asymmetry exists
among wings, and that
the Galaxy contains isolated chimneys and superbubbles releasing hot gas
into the halo in local structures. 

The outflow models predict a large associated column density of
\ion{O}{7}, since they contain large columns of gas at 10$^6$\,K and
$f$(\ion{O}{7}) at 10$^6$\,K is close to unity under CIE
conditions. We find that log\,$N$(\ion{O}{7})=15--16 in our vertical
outflow models. 
For comparison, \citet{Mc04} report five zero-redshift \ion{O}{7}
detections at 
log\,$N$ $>$16.10, $>$14.85, $>$15.85, $>$16.45, and $>$15.74, so the
model prediction (of a large reservoir of million-degree gas) is
consistent with X-ray observations. However, our observation that in 81\% of
cases \ct\ absorption accompanies the \hvo\ is
inconsistent with the \hvo\ existing at 10$^6$\,K in ionization
equilibrium with \ion{O}{7}.

The outflow model has the problem of explaining why the wings are
generally not seen in halo star spectra.
This problem could be avoided if the outflow were initially at
temperatures log\,$T>6.6$, too hot to form \os\ until it
reached higher $z$-distances. 
However, a higher initial temperature in
turn requires a higher density, otherwise the gas will not cool to
temperatures where \os\ can form and a wing will not be produced.
After running our models to investigate this, we
found we could {\it not} reproduce \os\ wings with high-temperature outflows.
The only way to explain the halo star non-detections of \hvo\ with an
outflow explanation is then by 
arguing that the \os\ absorbers are being missed against complicated
halo star continua, or by invoking a patchy, localized outflow.
The latter argument has trouble explaining why the \hvo\ patchiness should be 
different for halo star versus AGN spectra. 


The radial outflow model tends to produce narrow absorption
components, whereas the vertical model produces broader absorption
wings, closer to the observed \os\ wings. The 
radial model fails to reproduce this broad absorption wings because
projection 
effects result in the majority of the column density along a given sight
line being collapsed onto a narrow range in LSR velocity. 
The radial models require higher initial densities and velocities to
reproduce \os\ features at the right $v_{LSR}$; we found
$n_0\approx2\times10^{-2}$\,cm$^{-3}$ and $v_0\approx450$\kms\ was necessary.
A comparison
of data for four \hvo\ absorbers with the results of the outflow
modelling, showing the contrasting behavior of vertical and radial
models, is presented in Figure 14.
These models are all characterized by
$\dot{M}\approx10M_{\odot}\,$yr$^{-1}$ to each side of the Galactic
plane. This total flow rate would decrease if the covering factor of the
outflowing gas is $<$1.

The finding that radial outflows can produce \hvo\ components and vertical
outflows can produce \hvo\ wings results makes it very difficult to
interpret the data. Vertical and radial
outflows are unlikely to be simultaneous, so we do not consider that
outflows can explain all the highly ionized HVC observations. 
Therefore, while we have shown that the kinematics of the \os\ profiles can
broadly be reproduced by outflows, outflows are not demanded and we
cannot determine their geometry.

\subsection{Interfaces with low $N$(\hi) HVCs}
One explanation for the
highly ionized HVCs is that they are drawn from the same population as the
21\,cm-emitting \hi\ HVCs, being those with $N$(\hi)$<$10$^{18}$\sqcm.
This is suggested since both highly ionized and 21\,cm-bright HVCs
display a multi-phase structure with 
\hi\ and \os\ components. The distinction between \hi\ HVCs and highly
ionized HVCs becomes merely an \hi\ 21\,cm detection limit issue:
we detect 21\,cm emission only when
$N$(\hi)$\gtrsim$2$\times10^{18}$\sqcm, but we  
intercept \os-bearing interfaces at the front and back of
the \hi\ phase in all cases. 
This interpretation is supported by the observations that
the \hvo\ velocity centroids are correlated to the \hvh\ velocity
centroids (Fig. 10) and the
\hvo\ column densities (log\,$N\approx$13--14) 
remain the same in HVCs regardless of the presence of \hi\ 21\,cm
emission.
To achieve log\,$N$(\os) = 13.8, the observed average \os\ column
density in highly ionized HVCs, requires $\approx$6 conductive
interfaces, since each interface contributes
$N$(\os)$\approx10^{13}$\sqcm\ \citep{Bo90}. Therefore some cloud
fragmentation is needed to build up the observed
$N$(\os). Interfaces where the energy flow is turbulent 
rather than conductive are also possible \citep[e.g.][]{Es05}.

\subsection{Local Group vs Galactic Clouds}
The evidence from the halo star spectra implies the $|z|$-distance to the
\hvo\ absorbers is $\gtrsim10$\,kpc, and in the case
of the HE~0226-4110 absorbers, metallicity and kinematic measurements
suggest an association with the Magellanic Stream at
$\approx$50\,kpc \citep{Fo05}. However, in general, distance
information is lacking for the highly ionized HVCs. An alternative
explanation for the \hvo\ 
absorbers, which explains their overall kinematics, is that they form
a population of distant, Local Group clouds \citep{Ni03}.
While we cannot rule out the possibility that some highly ionized HVCs
exist in the
Local Group, we favor a Galactic origin for the following reasons:

{\it Mass}. \citet{Co05} consider the Local Group
explanation for highly ionized HVCs to be unlikely since the implied mass
($4\times10^{12}M_{\odot}$) it too large to be consistent with
gravitational models of the Local Group.

{\it Association with a Galactic corona}. 
\ion{O}{7} absorption has been detected toward a Galactic globular
cluster and toward the Large Magellanic Cloud, implying a hot, extended phase
of the Galactic interstellar medium exists \citep{Fu04,Wa05}. 
Our finding that over 80\% of highly ionized \hvo\ absorbers
show associated \ct\ and \hi\ strongly suggests
that the \hvo\ is not formed {\it in} the same gas as that producing
X-ray absorption in the lines of \ion{O}{7} and \ion{O}{8}. This
conclusion is the same as that reached by \citet{Wi05} and
\citet{Wi06} for the \hvo\ in the sight 
lines toward Mrk~421 and Mrk~279. However, much of the \hvo\ could
exist in interfaces between embedded clouds and the hotter plasma, so
the \hvo\ traces the corona indirectly. 

{\it Association with Galactic \hi\ HVCs}.
Evidence is becoming stronger that the \hi\ HVCs are a Galactic and
not an extragalactic phenomenon \citep[see, e.g.][]{Pi04,Oo04,Br05,Ri06}. 
Therefore, if \hi\ HVCs are Galactic, and highly ionized HVCs are simply
low $N$(\hi) HVCs as we have argued in this paper, then highly ionized
HVCs are Galactic by association. 

The \hi\ cores in the highly ionized HVCs could represent
cooled clumps of tidal 
material stripped off nearby galaxies (e.g. the Magellanic
Stream). Indeed, many highly ionized HVCs exist in the general
vicinity of the Magellanic Stream (\S4.1).
These cores will become photoionized since the low $N$(\hi)
prevents any self-shielding from the ambient radiation field \citep{Fo05}.
In the future, far-UV absorption line
studies could be used to extend the \hi\ HVC distribution function
to $N\ll10^{18}$\sqcm\ using the Lyman series absorption lines.


\section{Summary}
We summarize in the following points the empirical results of our
survey of \os, \ct, and \hi\ in HVCs using \fuse\ spectra of
extragalactic targets with S/N$_{1030}>$8. 

\begin{enumerate} 
\item  In a sample of 66 sight lines, 49 (74\%) show \hvo\
  absorption. 
  15 sight lines contain \os\ HVCs with \hi\ 21\,cm
  counterparts, and 34 sight lines contain ``highly ionized'' HVCs
  without \hi\ 
  21\,cm counterparts. With some sight lines containing multiple HVCs,
  47 highly ionized HVCs are seen in total (20 at $b>0$\degr\ and 27
  at $b<0$\degr).
 
\item Of the negative-velocity highly ionized HVCs, 82\% are at
  $l<180$\degr\ and 94\% are at $b<0$\degr. Of the positive-velocity
  highly ionized HVCs, 73\% are at $l>180$\degr\ and
  63\% are at $b>0$\degr. 

\item We classified the highly ionized HVCs into 11 \pvw s (broad \os\
  absorption line  
  features extending to velocities of 200--300\kms), 19 \pvc s
  (discrete absorbers), and 17 \nvc s. Over the whole sample, 
  $\langle b$(\os)$\rangle$=38$\pm$10\kms\ and
  $\langle$log\,$N$(\os)$\rangle$=13.83$\pm$0.36. 
  81\% of the \hvo\ absorbers have clear accompanying \ct\ absorption
  at the same 
  velocity, and 76\% of cases have accompanying \hi\
  absorption in the Lyman series. The cases where \hv\ \ct\ and \hvh\ are
  not detected are almost all weak, low column density \hvo\ absorbers.
  The likelihood of a \hvo\ absorber having a \ct\ or \hi\ counterpart
  is unconnected to position on the sky.

\item We present the first (\os-selected) sample of \ct\ and \hi\
  absorption line HVCs. For the 14 unsaturated \ct\ absorbers, we find 
  $\langle b$(\ct)$\rangle$=30$\pm$8\kms; log\,$N_a$(\ct) ranges from
  12.8 to $>$14.4. For the 11 unsaturated \hvh\ absorbers, we find
  $\langle b$(\hi)$\rangle$=22$\pm$5\kms; log\,$N_a$(\hi) ranges from
  15.1 to $>$16.9.

\item The upper limit on temperature provided by the average width of the
  high-velocity \hi\ lines is log\,$T<$4.67, where no \os\ is expected to
  exist. Therefore, the \os\ and \hi\
  exist in separate phases of gas. HVCs are multi-phase structures.

\item Photoionization is an unlikely explanation for the \hvo, since 
  the required cloud sizes are extremely large (S03) and inconsistent with
  those suggested by other ions \citep{Fo05}. Detailed comparisons of the
  \hvo\ and \ct\ line profiles
  indicate that $N_a$(\ct)/$N_a$(\os) is generally constant with
  velocity in highly ionized HVCs
  (with a value between 0 and 4), indicating that
  \os\ and \ct\ may coexist in the same, collisionally ionized phase. 
  The relative ion ratios of \os/\ct\ can be
  formed in CIE at temperatures near 1.7$\times10^5$\,K, assuming a
  solar elemental abundance pattern, though the high rate of cooling
  at these temperatures makes CIE unlikely. 

\item \Pvw s represent 23\% of all \hvo\ absorbers with
  no 21\,cm counterparts, and are seen in 14\% of all extragalactic
  sight lines observed with \fuse\ at S/N$>$8. 89\% of \os\ wing
  absorbers have counterparts in \ct\ and \hi. For the population of
  11 \pvw s in our sample, we measure 
  $\langle$log\,$N_a\rangle$=13.86$\pm$0.46,
  $\langle v_0\rangle$=134$\pm$28\kms, and 
  $\langle b\rangle$=40$\pm$11\kms. 
  8/11 wings are seen at $b>0$\degr, and 9/11 at $l>179$\degr, but
  wings are not seen at negative velocities. Wings are not seen
  toward halo stars out to $z\approx5$\,kpc, although continuum
  placement is difficult in these cases. The average properties of
  \os\ wings are statistically no different from \os\ components. 

\item The prevalence of \ct\ and \hi\ absorption
  kinematically associated with the \hvo\ rules out the idea that
  the \hvo\ is co-spatial with \ion{O}{7}- and
  \ion{O}{8}-absorbing gas. 


{\it We now discuss the implications of our results on the origins of
  the highly ionized HVCs.}

\item {\bf Galactic Outflows.} We produced one-dimensional ballistic
  models of cooling, decelerating Galactic 
  outflows to determine the \os\ line profiles expected in an outflow
  scenario. We find that vertical (fountain) outflows originating in
  the disk with $T_0=10^6$\,K, $n_0\approx2\times10^{-3}$\,cm$^{-3}$
  and $v_0\approx250$\kms\ can reproduce the kinematic 
  line profiles of the \os\ \pvw s. The radial outflows produce
  narrower absorption components, due to projection effects from our
  solar vantage point. In order to explain the
  observations, the Galactic outflow must be patchy (to deal with the
  wing non-detections toward other AGNs and halo stars) and out of CIE
  (to account for the \ct). We could not reproduce \os\ wings with
  higher temperature ($T_0>5\times10^6$\,K) outflows since the gas
  cooling time is too long.
  While the kinematics of the \os\
  profiles can broadly be reproduced by outflows, outflows are not
  demanded and we cannot determine their geometry from the current data.

\item {\bf Interfaces with low $N$(\hi) HVCs.} Our model of highly
  ionized HVCs (multi-phase structures containing an \hi\ core,
  a transition-temperature boundary layer, and a hot surrounding
  medium) is the same as that used to explain
  the structure of the large 21\,cm \hi-emitting complexes.
  It therefore seems reasonable to conclude that, with the possible exception
  of \pvw s, highly ionized HVCs represent the subset of HVCs having
  $N$(\hi)$\lesssim10^{18}$\sqcm, too small to be detected in 
  21\,cm emission. The similarity in the properties of \os\ absorption
  in 21\,cm-bright and highly ionized HVCs supports this model, as
  does the correlation between the \hvo\ velocity centroids and the
  \hvh\ velocity centroids.
%

\item {\bf Local Group vs Galactic Clouds.} We cannot rule out the
  possibility that some \hvo\ absorbers exist in the Local Group or
  beyond. However, we favor a Galactic origin based on the
  unreasonably high implied
  mass in the Local Group case \citep{Co05}, the recent evidence that
  \hi\ HVCs are Galactic \citep{Oo04} and that highly ionized HVCs are
  related to 
  the \hi\ HVCs (this paper), and the detection of a large reservoir of hot
  interstellar plasma around the galaxy \citep{Wa05} and that
  highly ionized HVCs can be formed in interfaces adjoining this plasma. 
  The non-detection of \hvo\ in halo star spectra implies 
  that any Galactic \hvo\ exists at $z$-distances beyond a few kpc.
\end{enumerate}

Acknowledgments\\
We thank Marilyn Meade for assisting with the CALFUSE data reduction
pipeline, and are grateful to the referee for a number of perceptive
comments. BDS 
acknowledges support through through NASA grant NNG04GK12G and the
University of Wisconsin Graduate School. BPW was supported by NASA
grants NAG5-9179 and NNG04GD85G.

{\it Facility:} \facility{FUSE}


\clearpage
\LongTables 
\begin{deluxetable}{lcclc ccc}
\tabletypesize{\tiny}
\tablewidth{0pt}
\tabcolsep=4.0pt
\tablecaption{All Sight Lines Searched for High-Velocity \os} 
\tablehead{Sight Line & $l$ & $b$ & Program ID &  CALFUSE & $t_{exp}$\tablenotemark{a} & $F_{1030}$\tablenotemark{b}
& S/N$_{1030}\tablenotemark{c}$  \\
& (\degr) & (\degr) & & & (ks) & (flux units) & }
\startdata
          \object{MRK279} &     115.04 &   \phs46.86 &               \dataset{P1080303} & v2.4 &     183.3 & 10.1 & 44.1 \\
    & & &     \dataset{P1080304} & v2.4   & &
        \\
    & & &     \dataset{D1540101} & v2.4   & &
        \\
          \object{MRK817} &     100.30 &   \phs53.50 &               \dataset{P1080403} & v2.1 &     187.4 &  9.6 & 41.5 \\
    & & &     \dataset{P1080404} & v2.1   & &
        \\
      \object{PG1259+593} &     120.60 &   \phs58.10 &               \dataset{P1080101} & v2.1 &     162.3 &  1.8 & 37.3 \\
    & & &     \dataset{P1080102} & v2.1   & &
        \\
    & & &     \dataset{P1080103} & v2.1   & &
        \\
    & & &     \dataset{P1080104} & v2.1   & &
        \\
    & & &     \dataset{P1080105} & v2.1   & &
        \\
    & & &     \dataset{P1080106} & v2.1   & &
        \\
    & & &     \dataset{P1080107} & v2.1   & &
        \\
    & & &     \dataset{P1080108} & v2.1   & &
        \\
    & & &     \dataset{P1080109} & v2.1   & &
        \\
          \object{MRK876} & \phn 98.30 &   \phs40.40 &               \dataset{P1073101} & v2.1 &     132.2 &  6.3 & 33.3 \\
    & & &     \dataset{D0280203} & v2.4   & &
        \\
         \object{NGC4151} &     155.10 &   \phs75.10 &               \dataset{C0920101} & v2.1 & \phn 96.8 &  7.3 & 32.9 \\
     \object{PKS2155-304} & \phn 17.73 &    $-$52.25 &               \dataset{P1080701} & v2.1 &     123.2 &  2.8 & 31.7 \\
    & & &     \dataset{P1080703} & v2.1   & &
        \\
    & & &     \dataset{P1080705} & v2.1   & &
        \\
         \object{3C273.0} &     290.00 &   \phs64.40 &               \dataset{P1013501} & v2.1 & \phn 42.3 &  6.9 & 31.2 \\
      \object{PG0804+761} &     138.28 &   \phs31.03 &               \dataset{P1011901} & v2.1 &     174.0 &  7.0 & 30.6 \\
    & & &     \dataset{P1011903} & v2.4   & &
        \\
    & & &     \dataset{S6011001} & v2.1   & &
        \\
    & & &     \dataset{S6011002} & v2.1   & &
        \\
          \object{MRK509} & \phn 36.00 &    $-$29.90 &               \dataset{X0170101} & v2.4 &     113.0 &  6.7 & 29.5 \\
    & & &     \dataset{X0170102} & v2.4   & &
        \\
    & & &     \dataset{P1080601} & v2.1   & &
        \\
         \object{NGC1068} &     172.10 &    $-$51.93 &               \dataset{P1110202} & v2.1 & \phn 22.6 &  2.3 & 29.3 \\
          \object{MRK421} &     179.80 &   \phs65.00 &               \dataset{P1012901} & v2.1 & \phn 83.8 &  9.6 & 29.1 \\
    & & &     \dataset{Z0100101} & v2.4   & &
        \\
    & & &     \dataset{Z0100102} & v2.4   & &
        \\
    & & &     \dataset{Z0100103} & v2.4   & &
        \\
       \object{H1821+643} & \phn 94.00 &   \phs27.42 &               \dataset{P1016402} & v2.4 &     280.2 &  3.0 & 27.9 \\
    & & &     \dataset{P1016405} & v2.4   & &
        \\
    & & &     \dataset{C0950201} & v2.4   & &
        \\
    & & &     \dataset{C0950202} & v2.4   & &
        \\
          \object{MRK335} &     108.76 &    $-$41.42 &               \dataset{P1010203} & v2.4 & \phn 97.0 &  7.1 & 27.0 \\
    & & &     \dataset{P1010204} & v2.1   & &
        \\
         \object{NGC1705} &     261.10 &    $-$38.70 &               \dataset{A0460102} & v2.4 & \phn 21.3 &  6.8 & 26.7 \\
    & & &     \dataset{A0460103} & v2.4   & &
        \\
      \object{PG1116+215} &     223.40 &   \phs68.20 &               \dataset{P1013101} & v2.1 & \phn 76.9 &  5.7 & 25.5 \\
    & & &     \dataset{P1013102} & v2.1   & &
        \\
    & & &     \dataset{P1013103} & v2.1   & &
        \\
    & & &     \dataset{P1013104} & v2.1   & &
        \\
    & & &     \dataset{P1013105} & v2.1   & &
        \\
     \object{HE0226-4110} &     253.90 &    $-$65.80 &               \dataset{P2071301} & v2.1 &     207.8 &  2.7 & 24.6 \\
    & & &     \dataset{P1019101} & v2.1   & &
        \\
    & & &     \dataset{P1019102} & v2.4   & &
        \\
    & & &     \dataset{P1019103} & v2.4   & &
        \\
    & & &     \dataset{P1019104} & v2.4   & &
        \\
    & & &     \dataset{D0270101} & v2.4   & &
        \\
    & & &     \dataset{D0270102} & v2.4   & &
        \\
    & & &     \dataset{D0270103} & v2.4   & &
        \\
      \object{PG0953+414} &     179.80 &   \phs51.70 &               \dataset{P1012201} & v2.1 & \phn 72.1 &  5.2 & 24.3 \\
    & & &     \dataset{P1012202} & v2.4   & &
        \\
         \object{MRK1383} &     349.20 &   \phs55.10 &               \dataset{P1014801} & v2.1 & \phn 64.5 &  6.6 & 23.1 \\
    & & &     \dataset{P2670101} & v2.1   & &
        \\
         \object{NGC5236} &     314.58 &   \phs31.97 &               \dataset{A0460505} & v2.4 & \phn 26.5 &  4.1 & 20.2 \\
        \object{TON S210} &     224.97 &    $-$83.16 &               \dataset{P1070301} & v2.1 & \phn 54.5 &  6.2 & 20.2 \\
    & & &     \dataset{P1070302} & v2.1   & &
        \\
      \object{PG0844+349} &     188.60 &   \phs38.00 &               \dataset{P1012002} & v2.4 & \phn 81.7 &  3.7 & 20.0 \\
    & & &     \dataset{D0280301} & v2.4   & &
        \\
    & & &     \dataset{D0280302} & v2.4   & &
        \\
    & & &     \dataset{D0280303} & v2.4   & &
        \\
    & & &     \dataset{D0280304} & v2.4   & &
        \\
      \object{PG1211+143} &     267.55 &   \phs74.31 &               \dataset{P1072001} & v2.4 & \phn 52.2 &  5.4 & 19.0 \\
     \object{PKS0558-504} &     258.00 &    $-$28.60 &               \dataset{P1011504} & v2.4 & \phn 93.4 &  3.3 & 18.8 \\
    & & &     \dataset{C1490601} & v2.4   & &
        \\
      \object{PKS0405-12} &     204.90 &    $-$41.80 &               \dataset{B0870101} & v2.1 &     140.5 &  2.2 & 18.6 \\
    & & &     \dataset{D1030101} & v2.4   & &
        \\
    & & &     \dataset{D1030102} & v2.4   & &
        \\
          \object{MRK290} & \phn 91.49 &   \phs47.95 &               \dataset{P1072901} & v2.1 & \phn 80.7 &  3.4 & 18.5 \\
    & & &     \dataset{D0760101} & v2.4   & &
        \\
    & & &     \dataset{D0760102} & v2.4   & &
        \\
    & & &     \dataset{E0840101} & v2.4   & &
        \\
        \object{VIIZw118} &     151.36 &   \phs25.99 &               \dataset{P1011604} & v2.1 &     165.4 &  2.0 & 18.3 \\
    & & &     \dataset{P1011605} & v2.4   & &
        \\
    & & &     \dataset{P1011606} & v2.4   & &
        \\
    & & &     \dataset{S6011301} & v2.1   & &
        \\
         \object{NGC3690} &     141.90 &   \phs55.41 &               \dataset{B0040201} & v2.1 & \phn 59.7 &  5.6 & 17.9 \\
    & & &     \dataset{B0040202} & v2.1   & &
        \\
     \object{PKS2005-489} &     350.40 &    $-$32.60 &               \dataset{P1073801} & v2.1 & \phn 49.2 &  5.0 & 17.5 \\
    & & &     \dataset{C1490301} & v2.1   & &
        \\
    & & &     \dataset{C1490302} & v2.1   & &
        \\
      \object{PG1011-040} &     246.50 &   \phs40.75 &               \dataset{B0790101} & v2.4 & \phn 86.3 &  2.6 & 17.2 \\
         \object{PHL1811} & \phn 47.47 &    $-$44.82 &               \dataset{P2071101} & v2.4 & \phn 47.7 &  4.9 & 17.1 \\
    & & &     \dataset{P1081001} & v2.4   & &
        \\
    & & &     \dataset{P1081002} & v2.4   & &
        \\
    & & &     \dataset{P1081003} & v2.4   & &
        \\
         \object{NGC4670} &     212.70 &   \phs88.60 &               \dataset{B0220301} & v2.1 & \phn 35.4 &  0.8 & 16.7 \\
    & & &     \dataset{B0220302} & v2.1   & &
        \\
    & & &     \dataset{B0220303} & v2.1   & &
        \\
         \object{MRK1513} & \phn 63.67 &    $-$29.07 &               \dataset{P1018301} & v2.4 & \phn 42.9 &  3.9 & 16.3 \\
    & & &     \dataset{P1018302} & v2.4   & &
        \\
    & & &     \dataset{P1018303} & v2.4   & &
        \\
         \object{NGC7469} & \phn 83.10 &    $-$45.47 &               \dataset{P1018703} & v2.4 & \phn 42.9 &  5.9 & 16.0 \\
      \object{ESO141-G55} &     338.20 &    $-$26.70 &               \dataset{I9040104} & v2.4 & \phn 40.4 &  5.2 & 16.0 \\
          \object{MRK205} &     125.45 &   \phs41.67 &               \dataset{Q1060203} & v2.4 &     206.3 &  1.3 & 15.9 \\
    & & &     \dataset{S6010801} & v2.1   & &
        \\
    & & &     \dataset{D0540101} & v2.4   & &
        \\
    & & &     \dataset{D0540102} & v2.4   & &
        \\
    & & &     \dataset{D0540103} & v2.4   & &
        \\
      \object{PG1302-102} &     308.60 &   \phs52.20 &               \dataset{P1080201} & v2.1 &     144.9 &  1.6 & 15.7 \\
    & & &     \dataset{P1080202} & v2.1   & &
        \\
    & & &     \dataset{P1080203} & v2.1   & &
        \\
          \object{MRK153} &     156.73 &   \phs56.01 &               \dataset{A0940101} & v2.4 & \phn 65.1 &  7.8 & 15.2 \\
   \object{MS0700.7+6338} &     152.50 &   \phs25.60 &               \dataset{P2072701} & v2.4 &     106.4 &  1.9 & 14.6 \\
    & & &     \dataset{S6011501} & v2.4   & &
        \\
    & & &     \dataset{D0550501} & v2.4   & &
        \\
      \object{PG1626+554} & \phn 84.51 &   \phs42.19 &               \dataset{C0370101} & v2.1 & \phn 91.2 &  1.5 & 14.1 \\
           \object{MRK59} &     111.54 &   \phs82.11 &               \dataset{A0360202} & v2.4 & \phn  9.8 &  2.2 & 13.9 \\
         \object{3C249.1} &     130.39 &   \phs38.55 &               \dataset{P1071601} & v2.1 &     219.4 &  1.1 & 13.8 \\
    & & &     \dataset{P1071602} & v2.1   & &
        \\
    & & &     \dataset{S6010901} & v2.1   & &
        \\
    & & &     \dataset{P1071603} & v2.4   & &
        \\
    & & &     \dataset{D1170101} & v2.4   & &
        \\
    & & &     \dataset{D1170102} & v2.4   & &
        \\
    & & &     \dataset{D1170103} & v2.4   & &
        \\
        \object{TON S180} &     139.00 &    $-$85.10 &               \dataset{P1010502} & v2.4 & \phn 28.6 &  6.3 & 13.6 \\
    & & &     \dataset{D0280101} & v2.4   & &
        \\
         \object{NGC7714} & \phn 88.22 &    $-$55.56 &               \dataset{A0230404} & v2.1 &     131.6 &  2.5 & 12.8 \\
    & & &     \dataset{A0860606} & v2.1   & &
        \\
    & & &     \dataset{B0040301} & v2.1   & &
        \\
    & & &     \dataset{C0370206} & v2.4   & &
        \\
    & & &     \dataset{C0370201} & v2.0   & &
        \\
    & & &     \dataset{C0370202} & v2.0   & &
        \\
    & & &     \dataset{C0370203} & v2.0   & &
        \\
    & & &     \dataset{C0370204} & v2.0   & &
        \\
    & & &     \dataset{C0370205} & v2.0   & &
        \\
         \object{NGC1741} &     203.72 &    $-$26.29 &               \dataset{C0480201} & v2.4 & \phn 29.3 &  5.2 & 12.7 \\
    & & &     \dataset{C0480202} & v2.4   & &
        \\
            \object{MRK9} &     158.36 &   \phs28.75 &               \dataset{P1071101} & v2.4 & \phn 61.5 &  2.3 & 12.6 \\
    & & &     \dataset{P1071102} & v2.4   & &
        \\
    & & &     \dataset{P1071103} & v2.4   & &
        \\
    & & &     \dataset{S6011601} & v2.1   & &
        \\
          \object{MRK106} &     161.14 &   \phs42.88 &               \dataset{C1490501} & v2.4 &     121.9 &  1.6 & 12.1 \\
          \object{MRK586} &     157.60 &    $-$54.93 &               \dataset{D0550101} & v2.4 & \phn 64.0 &  2.1 & 11.9 \\
    & & &     \dataset{D0550102} & v2.4   & &
        \\
      \object{PG1553+113} & \phn 21.91 &   \phs43.96 &               \dataset{E5260501} & v2.4 & \phn 50.7 &  2.7 & 11.9 \\
    & & &     \dataset{E5260502} & v2.4   & &
        \\
    & & &     \dataset{E5260503} & v2.4   & &
        \\
      \object{1H0717+714} &     144.00 &   \phs28.00 &               \dataset{Z9071301} & v2.4 & \phn 55.6 &  2.7 & 11.7 \\
      \object{1H0707-495} &     260.20 &    $-$17.70 &               \dataset{B1050101} & v2.4 & \phn 67.0 &  1.8 & 11.7 \\
    & & &     \dataset{B1050102} & v2.4   & &
        \\
    & & &     \dataset{B1050103} & v2.4   & &
        \\
     \object{MRC2251-178} & \phn 46.15 &    $-$61.32 &               \dataset{P1111010} & v2.1 & \phn 52.5 &  2.2 & 11.6 \\
      \object{ESO572-G34} &     286.10 &   \phs42.10 &               \dataset{B0220201} & v2.1 & \phn 25.5 &  6.4 & 11.2 \\
     \object{HS0624+6907} &     145.71 &   \phs23.35 &               \dataset{P1071001} & v2.1 &     116.0 &  1.0 & 11.0 \\
    & & &     \dataset{P1071002} & v2.1   & &
        \\
    & & &     \dataset{S6011201} & v2.1   & &
        \\
    & & &     \dataset{S6011202} & v2.1   & &
        \\
         \object{MRK1095} &     201.70 &    $-$21.13 &               \dataset{P1011201} & v2.1 & \phn 56.3 &  2.0 & 11.0 \\
    & & &     \dataset{P1011202} & v2.1   & &
        \\
    & & &     \dataset{P1011203} & v2.1   & &
        \\
          \object{NGC625} &     273.70 &    $-$73.10 &               \dataset{D0400101} & v2.4 & \phn 56.7 &  2.7 & 11.0 \\
  \object{IRAS08339+6517} &     150.45 &   \phs35.60 &               \dataset{B0040101} & v2.4 & \phn  0.0 &  2.4 & 10.7 \\
    & & &     \dataset{B0040102} & v2.4   & &
        \\
          \object{NGC985} &     180.84 &    $-$59.49 &               \dataset{P1010903} & v2.4 & \phn 68.0 &  3.3 & 10.7 \\
          \object{MRK477} & \phn 93.04 &   \phs56.82 &               \dataset{D1180101} & v2.4 &     157.7 &  1.0 & 10.2 \\
          \object{MRK501} & \phn 63.60 &   \phs38.86 &               \dataset{P1073301} & v2.1 & \phn 30.0 &  3.1 &  9.9 \\
    & & &     \dataset{C0810101} & v2.4   & &
        \\
         \object{NGC1399} &     236.72 &    $-$53.63 &               \dataset{A0880303} & v2.4 & \phn 28.3 &  4.7 &  9.1 \\
    & & &     \dataset{A0880304} & v2.4   & &
        \\
         \object{NGC1522} &     262.00 &    $-$45.97 &               \dataset{Z9090501} & v2.4 & \phn  1.0 &  4.8 &  9.0 \\
          \object{MRK829} & \phn 58.76 &   \phs63.25 &               \dataset{A0220401} & v2.1 & \phn 11.0 &  6.9 &  8.4 \\
          \object{MRK478} & \phn 59.20 &   \phs65.00 &               \dataset{P1110101} & v2.4 & \phn 25.2 &  3.5 &  8.4 \\
     \object{HE1143-1810} &     281.90 &   \phs41.70 &               \dataset{P1071901} & v2.1 & \phn  7.3 &  6.1 &  8.3 \\
      \object{ESO265-G23} &     285.91 &   \phs16.59 &               \dataset{A1210405} & v2.1 & \phn 48.7 &  1.2 &  7.9 \\
    & & &     \dataset{A1210407} & v2.1   & &
        \\
    & & &     \dataset{A1210408} & v2.1   & &
        \\
    & & &     \dataset{A1210409} & v2.1   & &
        \\
         \object{NGC5408} &     317.15 &   \phs19.50 &               \dataset{Z9090901} & v2.4 & \phn 16.9 &  5.4 &  7.5 \\
\enddata
\tablenotetext{a}{Combined exposure time.}
\tablenotetext{b}{Flux at 1030\,\AA; 1 flux unit = $10^{-14}$\,erg\sqcm\,s$^{-1}$\,\AA$^{-1}$.}
\tablenotetext{c}{S/N  per resolution element at 1030\,\AA\ in combined spectrum.}
\end{deluxetable}

\clearpage 
\LongTables %
\begin{landscape} 
\begin{deluxetable}{lc cccc cccc ccccc}
\tabletypesize{\tiny}
\tabcolsep=2.0pt
\tablecaption{Detections of Highly Ionized HVCs}
\tablehead{Sight Line & $v_{min,max}$\tablenotemark{a} & 
\multicolumn{4}{c}{\underline{\phm{aaaaaaaaaaaa}\os\phm{aaaaaaaaaaaa}}} & 
\multicolumn{4}{c}{\underline{\phm{aaaaaaaaaaaa}\ct\phm{aaaaaaaaaaaa}}} &
\multicolumn{5}{c}{\underline{\phm{aaaaaaaaaaaaaaaaaa}\hi\phm{aaaaaaaaaaaaaaaaaaa}}} \\
& & $\bar{v}$\tablenotemark{b} & $b$\tablenotemark{c} & log $N_a$\tablenotemark{d} & Sig.\tablenotemark{e} & 
    $\bar{v}$\tablenotemark{b} & $b$\tablenotemark{c} & log $N_a$\tablenotemark{d} & Sig.\tablenotemark{e} & 
 Line\tablenotemark{f} & $\bar{v}$\tablenotemark{b} & $b$\tablenotemark{c} & log $N_a$\tablenotemark{d} & Sig.\tablenotemark{e} }
\startdata
\multicolumn{10}{l}{Sight lines with high-negative-velocity \os\ absorption} \\
     \object{PKS2155-304} &   $-$300,  $-$200 &  $-$244$\pm$ 6 &  36$\pm$ 6 & 13.51$_{-0.11}^{+0.09}$ & 12.6 & 
       $-$271$\pm$ 4 &  30$\pm$ 4 & 13.49$_{-0.05}^{+0.04}$ & 18.8 & 
926 &  $-$253$\pm$ 2 &  18$\pm$ 6 & 15.5$_{-0.3}^{+0.3}$ & 10.3 \\
                          &   $-$200,  $-$ 70 &  $-$127$\pm$ 5 &  47$\pm$ 2 & 13.90$_{-0.05}^{+0.05}$ & 26.6 & 
       $-$159$\pm$ 3 &  44$\pm$ 2 & 13.91$_{-0.03}^{+0.03}$ & 41.6 & 
926 &  $-$126$\pm$ 5 &  32$\pm$ 6 & 16.1$_{-0.3}^{+0.3}$ & 28.9 \\
          \object{MRK335} &   $-$375,  $-$250 &  $-$302$\pm$ 5 &  45$\pm$ 2 & 13.95$_{-0.05}^{+0.05}$ & 18.8 & 
       $-$320$\pm$ 4 &  40$\pm$ 2 & 14.04$_{-0.07}^{+0.06}$ & 38.2 & 
926 & \nodata & \nodata & $>$16.4     &\nodata \\
                          &   $-$180,  $-$ 80 &  $-$137$\pm$10 &  41$\pm$ 2 & 13.74$_{-0.08}^{+0.07}$ & 14.0 & 
\nodata & \nodata & $>$14.24   & \nodata & 
926 & \nodata & \nodata & $>$16.2     &\nodata \\
        \object{TON S210} &   $-$270,  $-$125 &  $-$187$\pm$ 9 &  57$\pm$ 5 & 13.86$_{-0.09}^{+0.08}$ &  9.6 & 
\nodata & \nodata & $>$14.27   & \nodata & 
926 & \nodata & \nodata & $>$16.8     &\nodata \\
         \object{PHL1811} &   $-$365,  $-$250 &  $-$311$\pm$ 8 &  42$\pm$ 5 & 13.79$_{-0.12}^{+0.09}$ &  5.8 & 
\nodata &\nodata &\nodata\tablenotemark{g}&\nodata &
\nodata &\nodata &\nodata &\nodata \tablenotemark{g}&\nodata \\
                          &   $-$190,  $-$ 80 &  $-$140$\pm$ 4 &  33$\pm$ 3 & 14.31$_{-0.04}^{+0.03}$ & 17.5 & 
\nodata &\nodata &\nodata\tablenotemark{g}&\nodata &
\nodata &\nodata &\nodata &\nodata \tablenotemark{g}&\nodata \\
         \object{MRK1513} &   $-$390,  $-$220 &  $-$301$\pm$ 1 &  51$\pm$ 1 & 14.46$_{-0.04}^{+0.03}$ & 22.8 & 
\nodata & \nodata & $>$13.96   & \nodata & 
949 & \nodata & \nodata & $>$15.8     &\nodata \\
         \object{NGC7469} &   $-$380,  $-$250 &  $-$308$\pm$ 6 &  50$\pm$ 5 & 14.13$_{-0.06}^{+0.05}$ & 14.0 & 
\nodata &\nodata &\nodata\tablenotemark{g}&\nodata &
\nodata &\nodata &\nodata &\nodata \tablenotemark{g}&\nodata \\
                          &   $-$200,  $-$100 &  $-$169$\pm$ 5 &  31$\pm$ 2 & 14.04$_{-0.07}^{+0.06}$ & 12.8 & 
\nodata &\nodata &\nodata\tablenotemark{g}&\nodata &
\nodata &\nodata &\nodata &\nodata \tablenotemark{g}&\nodata \\
         \object{NGC7714} &   $-$310,  $-$220 &  $-$262$\pm$ 2 &  31$\pm$ 1 & 14.16$_{-0.06}^{+0.05}$ & 12.0 & 
\nodata & \nodata & $>$14.00   & \nodata & 
937 & \nodata & \nodata & $>$16.0     &\nodata \\
                          &   $-$200,  $-$100 &  $-$149$\pm$ 5 &  33$\pm$ 4 & 13.66$_{-0.20}^{+0.13}$ &  3.7 & 
\nodata & \nodata & $>$13.84   & \nodata & 
937 & \nodata & \nodata & $>$15.9     &\nodata \\
\tableline
\multicolumn{10}{l}{Sight lines with high-positive-velocity \os\ absorption} \\
          \object{MRK817} &  \phs 50, \phs110 & \phs 71$\pm$ 9 &  21$\pm$ 6 & 13.04$_{-0.40}^{+0.20}$ &  7.3 & 
      \phs 75$\pm$ 9 &  18$\pm$ 3 & 12.77$_{-0.15}^{+0.11}$ &  7.2 & 
926 & \phs 72$\pm$ 6 &  14$\pm$ 6 & 15.2$_{-0.3}^{+0.3}$ &  7.5 \\
                          &  \phs140, \phs220 & \phs189$\pm$ 5 &  21$\pm$ 4 & 13.03$_{-0.41}^{+0.21}$ &  6.1 & 
\nodata & \nodata & $<$12.54   & \nodata & 
926 & \nodata & \nodata & $<$15.1     &\nodata \\
         \object{NGC4151} &  \phs120, \phs240 & \phs170$\pm$ 5 &  37$\pm$ 4 & 13.96$_{-0.07}^{+0.06}$ &  6.8 & 
      \phs155$\pm$ 3 &  30$\pm$ 3 & 13.68$_{-0.05}^{+0.04}$ & 17.1 & 
972 & \phs158$\pm$ 3 &  25$\pm$ 6 & 15.2$_{-0.3}^{+0.3}$ & 34.2 \\
         \object{3C273.0} &  \phs170, \phs250 & \phs206$\pm$ 4 &  30$\pm$ 2 & 13.44$_{-0.13}^{+0.10}$ & 12.8 & 
\nodata &\nodata &\nodata\tablenotemark{g}&\nodata &
926 & \nodata & \nodata & $<$15.0     &\nodata \\
          \object{MRK421} &  \phs 65, \phs160 & \phs 94$\pm$ 8 &  30$\pm$ 7 & 13.48$_{-0.10}^{+0.08}$ &  8.2 & 
\nodata & \nodata & $<$12.84   & \nodata & 
949 & \nodata & \nodata & $<$14.7     &\nodata \\
         \object{NGC1705} &  \phs200, \phs400 & \phs300$\pm$ 4 &  62$\pm$ 4 & 14.33$_{-0.03}^{+0.03}$ & 27.4 & 
\nodata & \nodata & $>$14.40   & \nodata & 
926 & \nodata & \nodata & $>$16.8     &\nodata \\
      \object{PG1116+215} &  \phs105, \phs305 & \phs173$\pm$ 4 &  42$\pm$ 7 & 14.02$_{-0.08}^{+0.07}$ &  5.9 & 
\nodata & \nodata & $>$14.14   & \nodata & 
923 & \nodata & \nodata & $>$16.7     &\nodata \\
     \object{HE0226-4110} &  \phs100, \phs230 & \phs164$\pm$ 4 &  42$\pm$ 3 & 13.82$_{-0.07}^{+0.06}$ & 11.4 & 
\nodata & \nodata & $>$14.14   & \nodata & 
920 & \nodata & \nodata & $>$16.9     &\nodata \\
      \object{PG0953+414} &  \phs100, \phs225 & \phs137$\pm$14 &  47$\pm$ 7 & 13.49$_{-0.22}^{+0.15}$ &  4.7 & 
      \phs134$\pm$ 5 &  31$\pm$ 2 & 13.66$_{-0.06}^{+0.05}$ & 14.3 & 
923 & \phs122$\pm$ 6 &  21$\pm$ 6 & 16.1$_{-0.3}^{+0.3}$ & 12.6 \\
         \object{MRK1383} &  \phs100, \phs160 & \phs133$\pm$ 6 &  21$\pm$ 3 & 13.14$_{-0.34}^{+0.19}$ &  3.9 & 
      \phs116$\pm$12 &  24$\pm$ 4 & 13.14$_{-0.15}^{+0.11}$ &  5.3 & 
937 & \nodata & \nodata & $<$15.0     &\nodata \\
      \object{PG0844+349} &  \phs110, \phs230 & \phs148$\pm$ 8 &  45$\pm$ 4 & 13.63$_{-0.14}^{+0.10}$ &  5.4 & 
      \phs137$\pm$ 9 &  33$\pm$ 5 & 13.68$_{-0.08}^{+0.07}$ & 11.2 & 
926 & \phs128$\pm$ 9 &  25$\pm$ 6 & 15.8$_{-0.3}^{+0.3}$ &  7.3 \\
      \object{PG1211+143} &  \phs130, \phs220 & \phs173$\pm$ 9 &  35$\pm$ 3 & 13.34$_{-0.31}^{+0.18}$ &  3.3 & 
      \phs181$\pm$ 3 &  22$\pm$ 4 & 13.23$_{-0.10}^{+0.08}$ &  7.1 & 
930 & \phs175$\pm$ 2 &  23$\pm$ 6 & 15.7$_{-0.3}^{+0.3}$ & 10.5 \\
     \object{PKS0558-504} &  \phs210, \phs315 & \phs258$\pm$ 2 &  39$\pm$ 2 & 13.76$_{-0.11}^{+0.09}$ &  8.0 & 
      \phs260$\pm$20 &  27$\pm$ 3 & 13.67$_{-0.26}^{+0.16}$ &  8.1 & 
926 & \nodata & \nodata & $>$16.2     &\nodata \\
      \object{PKS0405-12} &  \phs110, \phs205 & \phs153$\pm$ 6 &  36$\pm$ 2 & 13.54$_{-0.18}^{+0.12}$ &  5.4 & 
      \phs139$\pm$ 6 &  41$\pm$ 5 & 13.35$_{-0.18}^{+0.13}$ &  3.9 & 
949 & \phs136$\pm$ 2 &  17$\pm$ 6 & 15.2$_{-0.3}^{+0.3}$ &  5.5 \\
     \object{PKS2005-489} &  \phs115, \phs220 & \phs143$\pm$ 8 &  30$\pm$ 4 & 13.71$_{-0.13}^{+0.10}$ &  6.7 & 
\nodata & \nodata & $>$13.69   & \nodata & 
926 & \phs166$\pm$ 5 &  26$\pm$ 6 & 16.1$_{-0.3}^{+0.3}$ & 14.5 \\
      \object{PG1011-040} &  \phs100, \phs240 & \phs144$\pm$ 7 &  46$\pm$ 3 & 14.18$_{-0.06}^{+0.05}$ & 14.2 & 
\nodata & \nodata & $>$14.06   & \nodata & 
949 & \nodata & \nodata & $>$16.1     &\nodata \\
                          &  \phs240, \phs320 & \phs279$\pm$20 &  29$\pm$ 2 & 13.53$_{-0.18}^{+0.13}$ &  4.8 & 
\nodata & \nodata & $<$13.07   & \nodata & 
949 & \phs263$\pm$ 9 &  21$\pm$ 6 & 15.1$_{-0.3}^{+0.3}$ &  6.5 \\
         \object{NGC4670} &  \phs320, \phs380 & \phs353$\pm$ 3 &  21$\pm$ 3 & 13.92$_{-0.06}^{+0.05}$ & 12.8 & 
      \phs347$\pm$ 3 &  21$\pm$ 3 & 13.34$_{-0.11}^{+0.09}$ &  7.0 & 
972 & \phs344$\pm$ 3 &  23$\pm$ 6 & 15.0$_{-0.3}^{+0.3}$ &  4.6 \\
      \object{ESO141-G55} &  \phs135, \phs210 & \phs174$\pm$ 4 &  25$\pm$ 4 & 13.35$_{-0.28}^{+0.17}$ &  3.3 & 
      \phs173$\pm$ 7 &  33$\pm$ 3 & 13.03$_{-0.22}^{+0.15}$ &  3.6 & 
926 & \nodata & \nodata & $<$15.2     &\nodata \\
      \object{PG1302-102} &  \phs190, \phs340 & \phs255$\pm$ 3 &  51$\pm$ 3 & 14.01$_{-0.09}^{+0.07}$ &  9.6 & 
\nodata & \nodata & $<$13.60   & \nodata & 
930 & \nodata & \nodata & $<$15.8     &\nodata \\
   \object{MS0700.7+6338} &  \phs 90, \phs180 & \phs134$\pm$ 3 &  29$\pm$ 2 & 13.68$_{-0.15}^{+0.11}$ &  5.0 & 
\nodata & \nodata & $<$13.49   & \nodata & 
949 & \nodata & \nodata & $<$15.5     &\nodata \\
      \object{1H0707-495} &  \phs 95, \phs195 & \phs143$\pm$ 3 &  28$\pm$ 3 & 13.68$_{-0.22}^{+0.14}$ &  3.9 & 
\nodata & \nodata & $>$13.70   & \nodata & 
926 & \nodata & \nodata & $>$16.4     &\nodata \\
      \object{ESO572-G34} &  \phs100, \phs275 & \phs166$\pm$ 6 &  61$\pm$ 4 & 14.42$_{-0.05}^{+0.04}$ & 15.3 & 
\nodata & \nodata & $>$14.16   & \nodata & 
930 & \nodata & \nodata & $>$16.6     &\nodata \\
          \object{NGC625} &  \phs115, \phs225 & \phs157$\pm$ 6 &  36$\pm$ 5 & 14.05$_{-0.11}^{+0.09}$ &  6.3 & 
\nodata &\nodata &\nodata\tablenotemark{g}&\nodata &
\nodata &\nodata &\nodata &\nodata \tablenotemark{g}&\nodata \\
     \object{HE1143-1810} &  \phs100, \phs200 & \phs140$\pm$ 7 &  40$\pm$ 2 & 14.20$_{-0.09}^{+0.08}$ &  9.6 & 
\nodata & \nodata & $>$14.21   & \nodata & 
937 & \nodata & \nodata & $>$16.5     &\nodata \\
                          &  \phs200, \phs300 & \phs246$\pm$ 7 &  32$\pm$ 4 & 13.80$_{-0.23}^{+0.15}$ &  3.7 & 
\nodata & \nodata & $<$13.37   & \nodata & 
937 & \nodata & \nodata & $<$15.5     &\nodata \\
\tableline
\multicolumn{10}{l}{Sight lines with high-negative- and high-positive-velocity \os\ absorption} \\
          \object{MRK509} &   $-$350,  $-$200 &  $-$260$\pm$ 3 &  45$\pm$ 3 & 14.15$_{-0.03}^{+0.03}$ & 27.1 & 
\nodata & \nodata & $>$14.39   & \nodata & 
926 & \nodata & \nodata & $>$16.7     &\nodata \\
                          &   $-$200,  $-$100 &  $-$151$\pm$ 7 &  42$\pm$ 2 & 13.80$_{-0.07}^{+0.06}$ & 15.5 & 
\nodata & \nodata & $>$13.84   & \nodata & 
926 & \nodata & \nodata & $>$16.0     &\nodata \\
                          &  \phs100, \phs175 & \phs123$\pm$10 &  28$\pm$ 3 & 13.61$_{-0.10}^{+0.08}$ & 11.1 & 
      \phs114$\pm$ 6 &  24$\pm$ 2 & 13.72$_{-0.06}^{+0.06}$ & 18.0 & 
926 & \nodata & \nodata & $>$16.1     &\nodata \\
        \object{TON S180} &   $-$200,  $-$100 &  $-$151$\pm$ 4 &  34$\pm$ 2 & 14.20$_{-0.05}^{+0.05}$ & 15.9 & 
\nodata & \nodata & $>$14.22   & \nodata & 
972 & \nodata & \nodata & $>$15.8     &\nodata \\
                          &  \phs215, \phs290 & \phs251$\pm$ 2 &  22$\pm$ 3 & 13.66$_{-0.15}^{+0.11}$ &  5.4 & 
\nodata & \nodata & $<$13.31   & \nodata & 
972 & \nodata & \nodata & $<$14.7     &\nodata \\
         \object{NGC1522} &   $-$250,  $-$100 &  $-$178$\pm$ 4 &  54$\pm$ 3 & 13.86$_{-0.22}^{+0.15}$ &  3.2 & 
\nodata &\nodata &\nodata\tablenotemark{g}&\nodata &
\nodata &\nodata &\nodata &\nodata \tablenotemark{g}&\nodata \\
                          &  \phs205, \phs360 & \phs284$\pm$ 3 &  51$\pm$ 3 & 14.15$_{-0.09}^{+0.07}$ &  7.6 & 
\nodata &\nodata &\nodata\tablenotemark{g}&\nodata &
\nodata &\nodata &\nodata &\nodata \tablenotemark{g}&\nodata \\
         \object{NGC5408} &   $-$200,  $-$100 &  $-$151$\pm$ 5 &  38$\pm$ 3 & 13.96$_{-0.16}^{+0.12}$ &  4.8 & 
\nodata &\nodata &\nodata\tablenotemark{g}&\nodata &
\nodata &\nodata &\nodata &\nodata \tablenotemark{g}&\nodata \\
                          &  \phs100, \phs250 & \phs148$\pm$ 8 &  51$\pm$ 3 & 14.50$_{-0.07}^{+0.06}$ & 11.4 & 
\nodata &\nodata &\nodata\tablenotemark{g}&\nodata &
\nodata &\nodata &\nodata &\nodata \tablenotemark{g}&\nodata \\
\enddata
\tablenotetext{a}{$v_{min}$ and $v_{max}$ are the observed minimum and maximum velocity of \hvo\ absorption.}
\tablenotetext{b}{Velocity centroid of high-velocity absorption, obtained from
$\bar{v}=\int^{v_{max}}_{v_{min}}v\tau_a(v)\mathrm{d}v/\int^{v_{max}}_{v_{min}}\tau_a(v)\mathrm{d}v$.} 
\tablenotetext{c}{Doppler parameter (width) of high-velocity
absorption, obtained from
$b=\sqrt{2\int^{v_{max}}_{v_{min}}(v-\bar{v})^2\tau_a(v)\mathrm{d}v/\int^{v_{max}}_{v_{min}}\tau_a(v)\mathrm{d}v}$.}
\tablenotetext{d}{Column densities calculated using the apparent optical depth
(AOD) technique, including statistical, continuum
placement, and systematic errors.}
\tablenotetext{e}{Detection significance, $W_{\lambda}/\sigma(W_{\lambda})$.}
\tablenotetext{f}{\hi\ Lyman series line used in measurement.}
\tablenotetext{g}{No measurement possible, due to blending or insufficient flux in SiC channels.}
\end{deluxetable}

\clearpage 
\end{landscape} 
\begin{deluxetable}{lcccc}
\tabletypesize{\tiny}
\tablewidth{0pt}
\tabcolsep=4.0pt
\tablecaption{3$\sigma$ Non-Detections of Highly Ionized HVCs}
\tablehead{Sight Line & 
\multicolumn{2}{c}{\underline{\phm{zzz}$-$200 to $-$100\kms\phm{zzz}}} & 
\multicolumn{2}{c}{\underline{\phm{zzz}100 to 200\kms\phm{zzz}}} \\
& $W_{\lambda}$ & log\,$N$(\os) &
$W_{\lambda}$ & log\,$N$(\os) \\
& (m\AA) & ($N$ in \sqcm) &
(m\AA) & ($N$ in \sqcm)}
\startdata
      \object{PG1553+113} & 
$<$ 80 & $<$13.81 & 
$<$ 21 & $<$13.24 \\
     \object{MRC2251-178} & 
$<$ 67 & $<$13.73 & 
$<$ 46 & $<$13.57 \\
          \object{MRK829} & 
$<$119 & $<$13.98 & 
$<$100 & $<$13.91 \\
          \object{MRK478} & 
$<$122 & $<$13.99 & 
$<$ 59 & $<$13.68 \\
          \object{MRK477} & 
$<$102 & $<$13.92 & 
$<$ 97 & $<$13.89 \\
           \object{MRK59} & 
$<$ 51 & $<$13.61 & 
$<$ 36 & $<$13.46 \\
      \object{1H0717+714} & 
$<$ 82 & $<$13.82 & 
$<$ 46 & $<$13.57 \\
        \object{VIIZw118} & 
$<$ 37 & $<$13.48 & 
$<$ 38 & $<$13.49 \\
          \object{MRK153} & 
$<$ 14 & $<$13.07 & 
$<$ 68 & $<$13.74 \\
          \object{MRK586} & 
$<$ 67 & $<$13.73 & 
$<$ 20 & $<$13.21 \\
            \object{MRK9} & 
$<$ 43 & $<$13.54 & 
$<$ 56 & $<$13.66 \\
          \object{MRK106} & 
$<$ 71 & $<$13.76 & 
$<$ 37 & $<$13.48 \\
         \object{NGC1068} & 
$<$ 36 & $<$13.47 & 
$<$ 12 & $<$12.99 \\
          \object{NGC985} & 
$<$ 72 & $<$13.76 & 
$<$ 32 & $<$13.41 \\
         \object{MRK1095} & 
$<$ 50 & $<$13.60 & 
$<$  5 & $<$12.62 \\
         \object{NGC1741} & 
$<$ 67 & $<$13.73 & 
$<$ 57 & $<$13.66 \\
         \object{NGC1399} & 
$<$157 & $<$14.10 & 
$<$ 19 & $<$13.20 \\
\tableline
     \object{PKS2155-304} & 
             \nodata & \nodata\tablenotemark{a} & $<$21 & $<$13.23 \\
        \object{TON S210} & 
             \nodata & \nodata\tablenotemark{a} & $<$32 & $<$13.42 \\
         \object{PHL1811} & 
             \nodata & \nodata\tablenotemark{a} & $<$58 & $<$13.67 \\
         \object{MRK1513} & 
             \nodata & \nodata\tablenotemark{a} & $<$34 & $<$13.44 \\
         \object{NGC7469} & 
             \nodata & \nodata\tablenotemark{a} & $<$24 & $<$13.30 \\
         \object{NGC7714} & 
             \nodata & \nodata\tablenotemark{a} & $<$43 & $<$13.54 \\
         \object{NGC4151} & 
$<$18 & $<$13.18           & \nodata & \nodata\tablenotemark{a} \\
         \object{3C273.0} & 
$<$30 & $<$13.38           & \nodata & \nodata\tablenotemark{a} \\
          \object{MRK421} & 
$<$27 & $<$13.34           & \nodata & \nodata\tablenotemark{a} \\
         \object{NGC1705} & 
$<$14 & $<$13.07           & \nodata & \nodata\tablenotemark{a} \\
      \object{PG1116+215} & 
$<$18 & $<$13.17           & \nodata & \nodata\tablenotemark{a} \\
     \object{HE0226-4110} & 
$<$12 & $<$13.00           & \nodata & \nodata\tablenotemark{a} \\
      \object{PG0953+414} & 
$<$49 & $<$13.60           & \nodata & \nodata\tablenotemark{a} \\
         \object{MRK1383} & 
$<$23 & $<$13.27           & \nodata & \nodata\tablenotemark{a} \\
      \object{PG0844+349} & 
$<$37 & $<$13.47           & \nodata & \nodata\tablenotemark{a} \\
      \object{PG1211+143} & 
$<$51 & $<$13.61           & \nodata & \nodata\tablenotemark{a} \\
     \object{PKS0558-504} & 
$<$44 & $<$13.55           & \nodata & \nodata\tablenotemark{a} \\
      \object{PKS0405-12} & 
$<$26 & $<$13.33           & \nodata & \nodata\tablenotemark{a} \\
     \object{PKS2005-489} & 
$<$52 & $<$13.63           & \nodata & \nodata\tablenotemark{a} \\
      \object{PG1011-040} & 
$<$32 & $<$13.41           & \nodata & \nodata\tablenotemark{a} \\
         \object{NGC4670} & 
$<$23 & $<$13.28           & \nodata & \nodata\tablenotemark{a} \\
      \object{ESO141-G55} & 
$<$51 & $<$13.61           & \nodata & \nodata\tablenotemark{a} \\
      \object{PG1302-102} & 
$<$46 & $<$13.57           & \nodata & \nodata\tablenotemark{a} \\
   \object{MS0700.7+6338} & 
$<$48 & $<$13.59           & \nodata & \nodata\tablenotemark{a} \\
      \object{1H0707-495} & 
$<$33 & $<$13.43           & \nodata & \nodata\tablenotemark{a} \\
      \object{ESO572-G34} & 
$<$98 & $<$13.89           & \nodata & \nodata\tablenotemark{a} \\
          \object{NGC625} & 
$<$78 & $<$13.80           & \nodata & \nodata\tablenotemark{a} \\
     \object{HE1143-1810} & 
$<$77 & $<$13.79           & \nodata & \nodata\tablenotemark{a} \\
\tableline
                   \object{MRK501} (ComplexC) & 
             \nodata & \nodata\tablenotemark{b} & $<$63 & $<$13.70 \\
               \object{PG1626+554} (ComplexC) & 
             \nodata & \nodata\tablenotemark{b} & $<$45 & $<$13.56 \\
                   \object{MRK290} (ComplexC) & 
             \nodata & \nodata\tablenotemark{b} & $<$41 & $<$13.52 \\
                \object{H1821+643} (OuterArm) & 
             \nodata & \nodata\tablenotemark{b} & $<$10 & $<$12.94 \\
                   \object{MRK876} (ComplexC) & 
             \nodata & \nodata\tablenotemark{b} & $<$23 & $<$13.27 \\
                   \object{MRK279} (ComplexC) & 
             \nodata & \nodata\tablenotemark{b} & $<$12 & $<$13.01 \\
               \object{PG1259+593} (ComplexC) & 
             \nodata & \nodata\tablenotemark{b} & $<$24 & $<$13.29 \\
                   \object{MRK205} (ComplexC) & 
             \nodata & \nodata\tablenotemark{b} & $<$38 & $<$13.49 \\
              \object{3C249.1} (nearComplexC) & 
             \nodata & \nodata\tablenotemark{b} & $<$35 & $<$13.46 \\
           \object{PG0804+761} (nearComplexA) & 
             \nodata & \nodata\tablenotemark{b} & $<$10 & $<$12.93 \\
              \object{NGC3690} (nearComplexC) & 
             \nodata & \nodata\tablenotemark{b} & $<$14 & $<$13.05 \\
              \object{HS0624+6907} (OuterArm) & 
             \nodata & \nodata\tablenotemark{b} & $<$62 & $<$13.70 \\
           \object{IRAS08339+6517} (ComplexA) & 
             \nodata & \nodata\tablenotemark{b} & $<$74 & $<$13.78 \\
              \object{ESO265-G23} (MagStream) & 
$<$71 & $<$13.76              & \nodata & \nodata\tablenotemark{b}
 \\
                 \object{NGC5236} (MagStream) & 
$<$80 & $<$13.81              & \nodata & \nodata\tablenotemark{b}
 \\
\enddata
\tablenotetext{a}{See Table 2 for detections in this velocity range.}
\tablenotetext{b}{Velocity range excluded due to the presence of a 21\,cm emitting \hi\ HVC, named in parentheses after the target name.}
\end{deluxetable}

\clearpage
\begin{deluxetable}{lcccc ccc}
\tablewidth{0pt}
\tabletypesize{\footnotesize}
\tablecaption{Analysis of Highly Ionized HVCs}
\tablehead{
&(1)&(2)&(3)&(4)&(5)&(6)&(7)\\
& $b<0$\degr & $b>0$\degr & $l<180$\degr &
 $l>180$\degr & With \ct\ & With \hi & Wings}
\startdata
       All Absorbers &  57\% (27/47) &  43\% (20/47) &  47\% (22/47) &  53\% (25/47) &  81\% (30/37) &  76\% (29/38) &  23\% (11/47   )
                     \\
           With \ct\ &  63\% (19/30) &  37\% (11/30) &  47\% (14/30) &  53\% (16/30) & \nodata &  93\% (28/30) &  27\% ( 8/30   )
                     \\
           With \hi\ &  62\% (18/29) &  38\% (11/29) &  48\% (14/29) &  52\% (15/29) &  97\% (28/29) & \nodata &  28\% ( 8/29   )
                     \\
               Wings &  27\% ( 3/11) &  73\% ( 8/11) &  36\% ( 4/11) &  64\% ( 7/11) &  89\% ( 8/ 9) &  89\% ( 8/ 9   ) & \nodata
                     \\
          $v_{LSR} < 0$\kms &  94\% (16/17) &   6\% ( 1/17) &  82\% (14/17) &  18\% ( 3/17) &  65\% (11/17) &  65\% (11/17) &   0\% ( 0/17   )
                            \\
          $v_{LSR} > 0$\kms &  37\% (11/30) &  63\% (19/30) &  27\% ( 8/30) &  73\% (22/30) &  63\% (19/30) &  60\% (18/30) &  37\% (11/30   )
                            \\
\enddata
\tablecomments{Each row refers to a sub-category of the highly ionized HVCs in this study; the table entries in each column represent the percentage of cases in that sub-category falling in different parts of the sky (columns 1 to 4), showing accompanying \ct\ absorption (column 5), showing accompanying \hi\ absorption (column 6), or appearing in the form of wings (column 7).}
\end{deluxetable}

\begin{deluxetable}{lcccc cccc}
\tablewidth{0pt}
\tabletypesize{\footnotesize}
\tablecaption{Properties of Wings vs Components\tablenotemark{a}}
\tablehead{Category & Num. & $\langle$log\,$N$(\os)$\rangle$ & 
$\langle v_0$(\os)$\rangle$ & $\langle b$(\os)$\rangle$ & With \ct\ & 
$\protect\left\langle\frac{N_a(\mathrm{C~III})}{N_a(\mathrm{O~VI})}\right\rangle$\tablenotemark{b}
& With \hi\ & $\left\langle\frac{N_a(\mathrm{H~I})}{N_a(\mathrm{O~VI})}\right\rangle$\tablenotemark{b} \\
& & ($N$ in cm$^{-2})$ & (km\,s$^{-1}$) & (km\,s$^{-1}$) & & & & }
\startdata
        Wings & 11 & 13.85$\pm$0.45   &\phs133$\pm$28 &   39$\pm$11 & 88\% & 1.1$\pm$0.4 & 88\% & 260$\pm$140 \\
PV Components & 19 & 13.69$\pm$0.34   &\phs212$\pm$63 &   34$\pm$11 & 64\% & 0.6$\pm$0.2 & 55\% &  70$\pm$100 \\
NV Components & 17 & 13.97$\pm$0.24 & $-$207$\pm$69 &   41$\pm$ 8 & 100\% & 1.1$\pm$0.1 & 100\% & 130$\pm$ 50 \\
\enddata
\tablenotetext{a}{Errors quoted on average quantities represent standard deviation of sample.}
\tablenotetext{b}{Only detections are included in this calculation. Cases with upper/lower limits on $N$(\ct) and $N$(\hi) are ignored.}
\end{deluxetable}

\clearpage
\begin{deluxetable}{lccc}
\tablewidth{0pt}
\tabletypesize{\scriptsize}
\tablecaption{Ionic Ratios and CIE Temperatures\tablenotemark{a}}
\tablehead{Target & Type\tablenotemark{b} & 
$N_a$(\ct)/$N_a$(\os) & $T_{CIE}$ }
\startdata
          \object{PKS2155-304} & NVC & 1.0$\pm$0.2 & 5.23 \\
                               & NVC & 1.0$\pm$0.1 & 5.23 \\
               \object{MRK335} & NVC & 1.2$\pm$0.2 & 5.22 \\
                               & NVC & $>$3.2 & $<$5.21 \\
             \object{TON S210} & NVC & $>$2.6 & $<$5.21 \\
              \object{MRK1513} & NVC & $>$0.3 & $<$5.25 \\
              \object{NGC7714} & NVC & $>$0.7 & $<$5.23 \\
                               & NVC & $>$1.5 & $<$5.22 \\
               \object{MRK817} & PVW & 0.5$\pm$0.4 & 5.24 \\
                               & PVC & $<$0.3 & $>$5.25 \\
              \object{NGC4151} & PVC & 0.5$\pm$0.1 & 5.24 \\
               \object{MRK421} & PVW & $<$0.2 & $>$5.25 \\
              \object{NGC1705} & PVC & $>$1.2 & $<$5.22 \\
           \object{PG1116+215} & PVC & $>$1.3 & $<$5.22 \\
          \object{HE0226-4110} & PVC & $>$2.1 & $<$5.22 \\
           \object{PG0953+414} & PVW & 1.5$\pm$0.6 & 5.22 \\
              \object{MRK1383} & PVC & 1.0$\pm$0.6 & 5.23 \\
           \object{PG0844+349} & PVW & 1.1$\pm$0.4 & 5.23 \\
           \object{PG1211+143} & PVC & 0.8$\pm$0.4 & 5.23 \\
          \object{PKS0558-504} & PVC & 0.8$\pm$0.4 & 5.23 \\
           \object{PKS0405-12} & PVC & 0.6$\pm$0.3 & 5.24 \\
          \object{PKS2005-489} & PVW & $>$1.0 & $<$5.23 \\
           \object{PG1011-040} & PVW & $>$0.8 & $<$5.23 \\
                               & PVC & $<$0.3 & $>$5.25 \\
              \object{NGC4670} & PVC & 0.3$\pm$0.1 & 5.25 \\
           \object{ESO141-G55} & PVC & 0.5$\pm$0.3 & 5.24 \\
           \object{PG1302-102} & PVC & $<$0.4 & $>$5.24 \\
        \object{MS0700.7+6338} & PVC & $<$0.6 & $>$5.24 \\
           \object{1H0707-495} & PVC & $>$1.0 & $<$5.23 \\
           \object{ESO572-G34} & PVW & $>$0.5 & $<$5.24 \\
               \object{NGC625} & PVW & $<$0.6 & $>$5.24 \\
          \object{HE1143-1810} & PVW & $>$1.0 & $<$5.23 \\
                               & PVC & $<$0.4 & $>$5.24 \\
               \object{MRK509} & NVC & $>$1.7 & $<$5.22 \\
                               & NVC & $>$1.1 & $<$5.23 \\
               \object{MRK509} & PVW & 1.3$\pm$0.3 & 5.22 \\
             \object{TON S180} & NVC & $>$1.0 & $<$5.23 \\
                               & PVC & $<$0.4 & $>$5.24 \\
\enddata
\tablenotetext{a}{Gas temperatures assume collisional ionization
equilibrium and solar carbon and oxygen abundances.}
\tablenotetext{b}{Type: PVC=\pvc, PVW=\pvw, NVC=\nvc. Multiple components in a given sight line are presented in order of increasing velocity.}
\end{deluxetable}

\clearpage
\begin{figure}[ht]
\epsscale{0.4}
\plotone{f1a.eps}
\plotone{f1b.eps}
\caption{Distribution of negative (left) and positive (right) highly
  ionized HVCs on the sky, using Aitoff 
  projections centered at $l=180$\degr, with contours showing \hi\
  21\,cm-emitting structures \citep[Complexes A and C, the anti-center
  clouds, and the Magellanic Stream;][]{HW88}. 
  Circles represent detections, coded by central velocity 
  (blue: $-400<v<-150$\kms; green: $-150<v<-100$\kms; 
  orange: $100<v<150$\kms; red: $150<v<400$\kms) and triangles show
  non-detections. Symbol sizes are proportional to the column density in the
  \os\ absorber (or to the upper limit for non-detections).}
\end{figure}

\clearpage
\begin{figure}[ht]
\epsscale{0.95}
\plotone{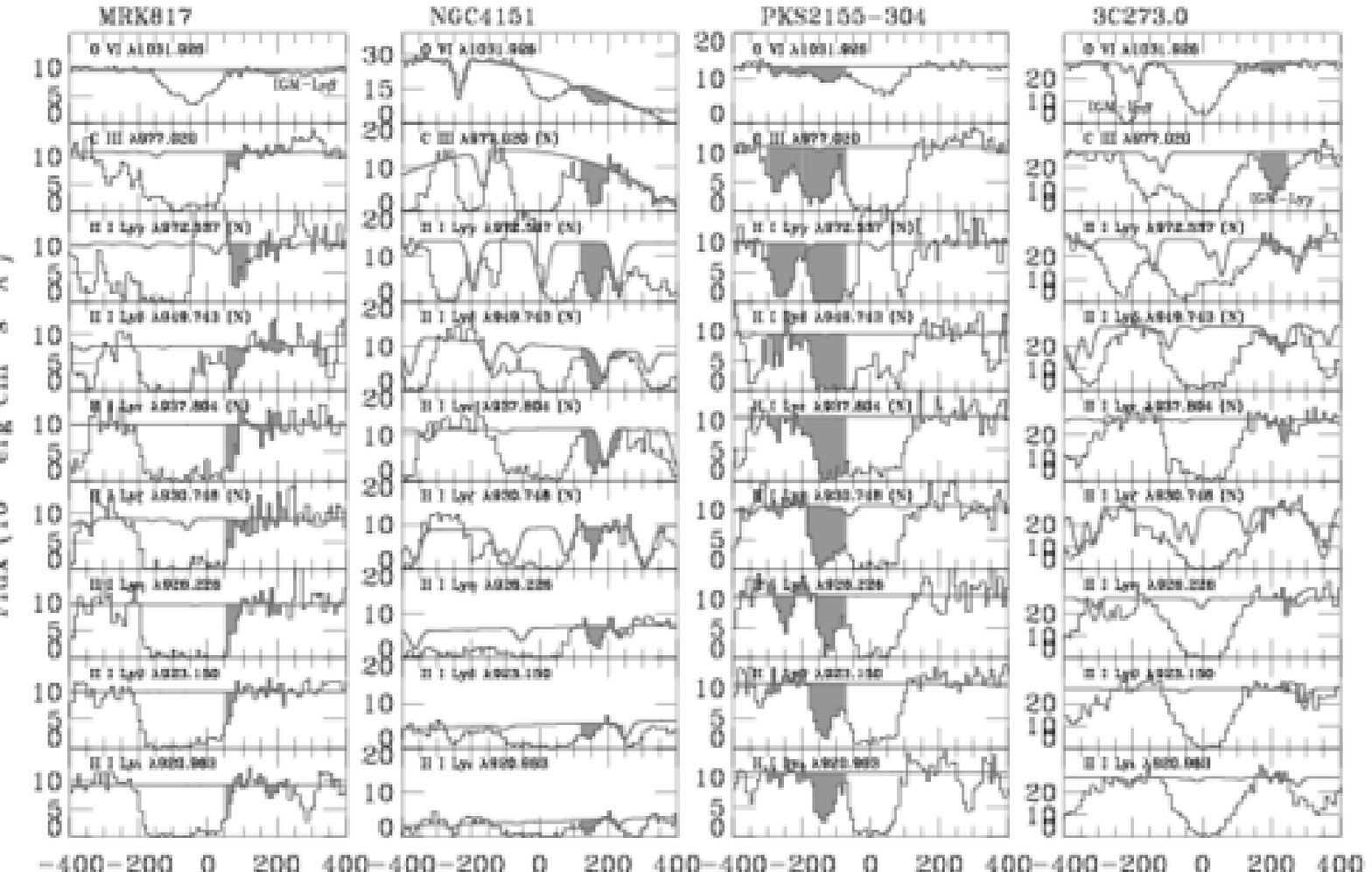}
\plotone{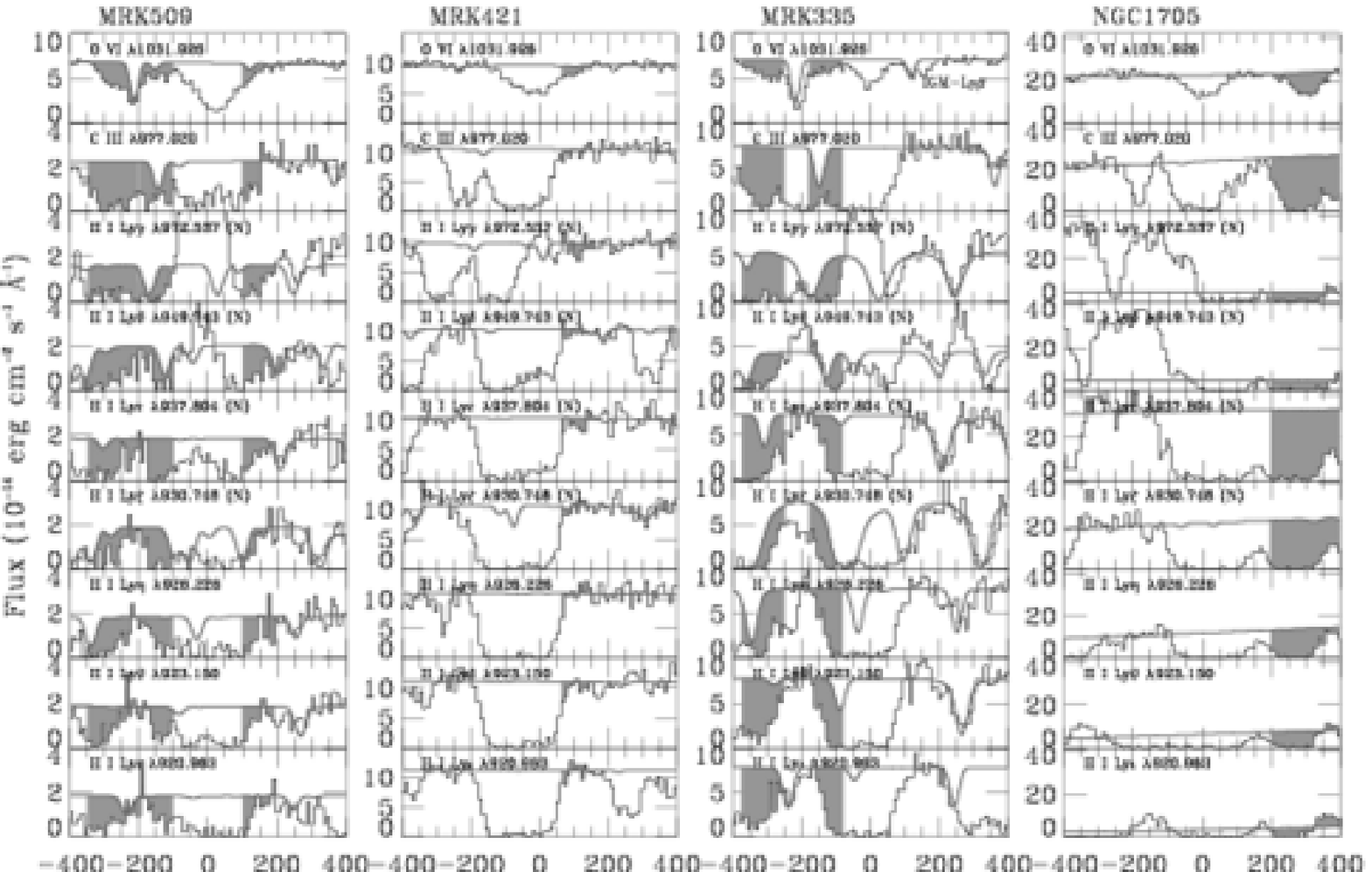}
\caption{\os, \ct, and \hi\ spectra for all highly ionized HVCs,
  presented in order of decreasing S/N. 
  Gray shading denotes velocity ranges where \hva\ is
  seen. The continuum placement, 
  including a model for H$_2$ model, is shown as a gray solid line in
  each panel.}
\end{figure}

\begin{figure}[ht]
\epsscale{1.0}
\figurenum{2 cont}
\plotone{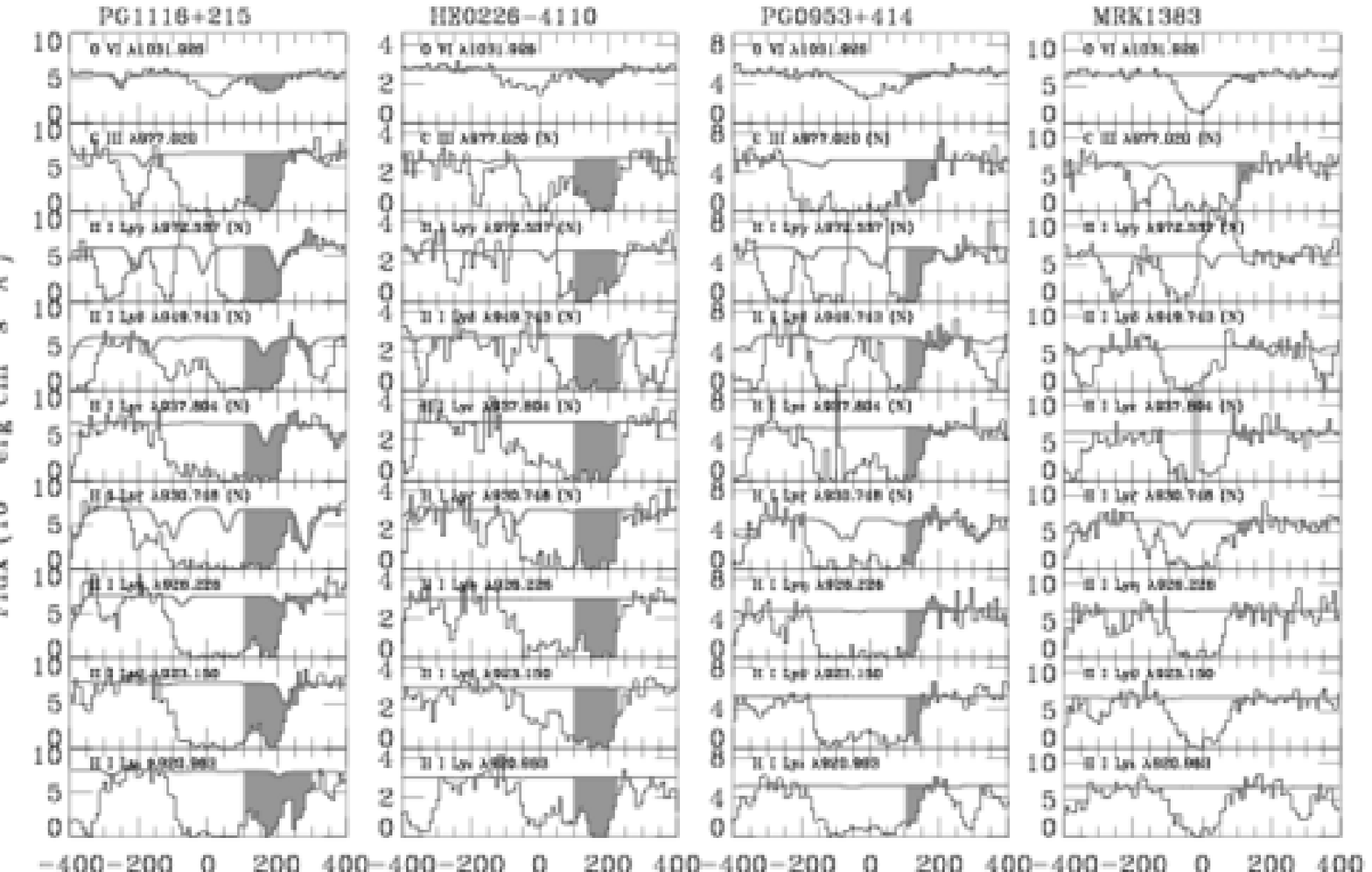}
\plotone{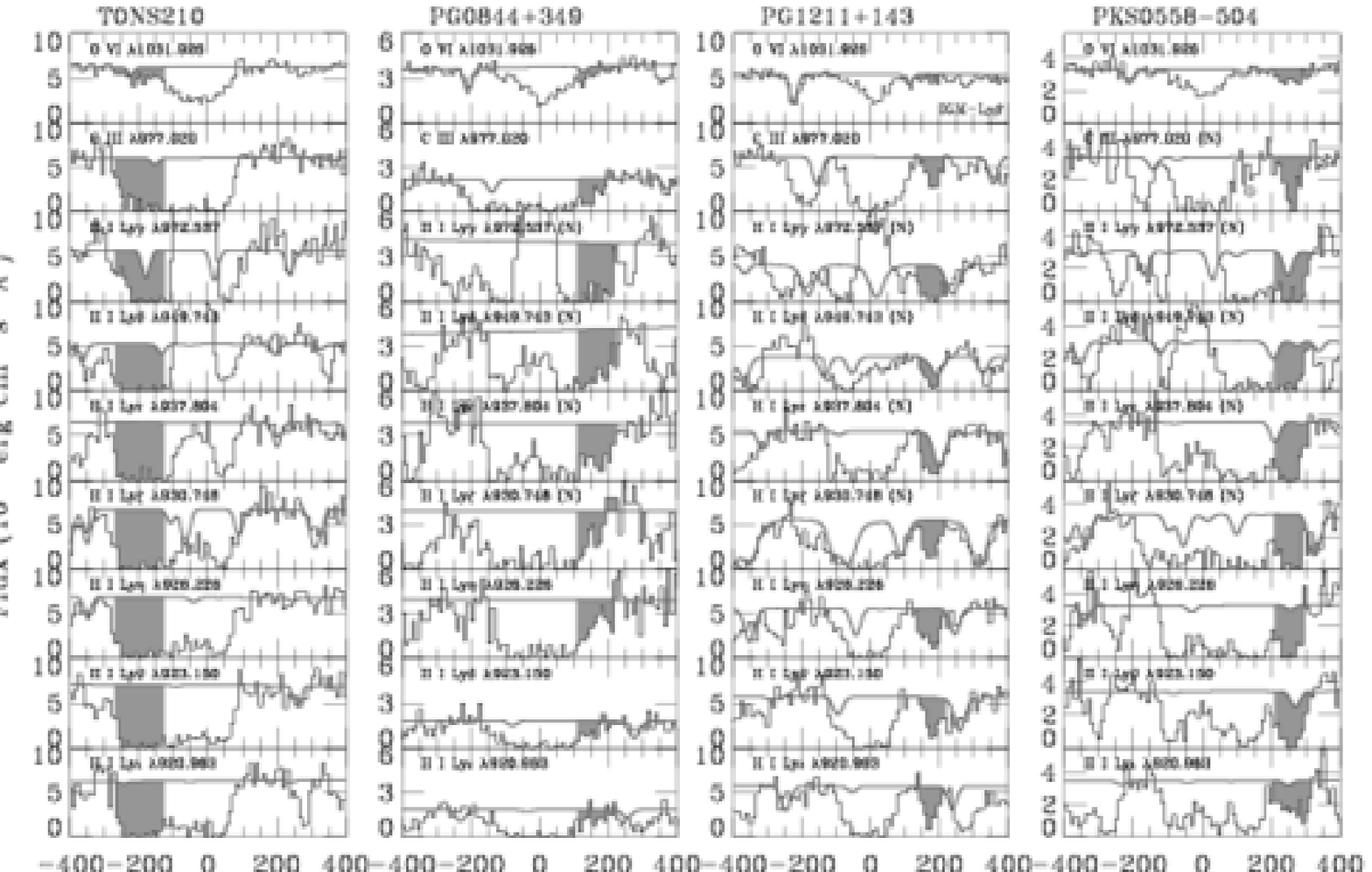}
\caption{}
\end{figure}

\begin{figure}[ht]
\epsscale{1.0}
\figurenum{2 cont}
\plotone{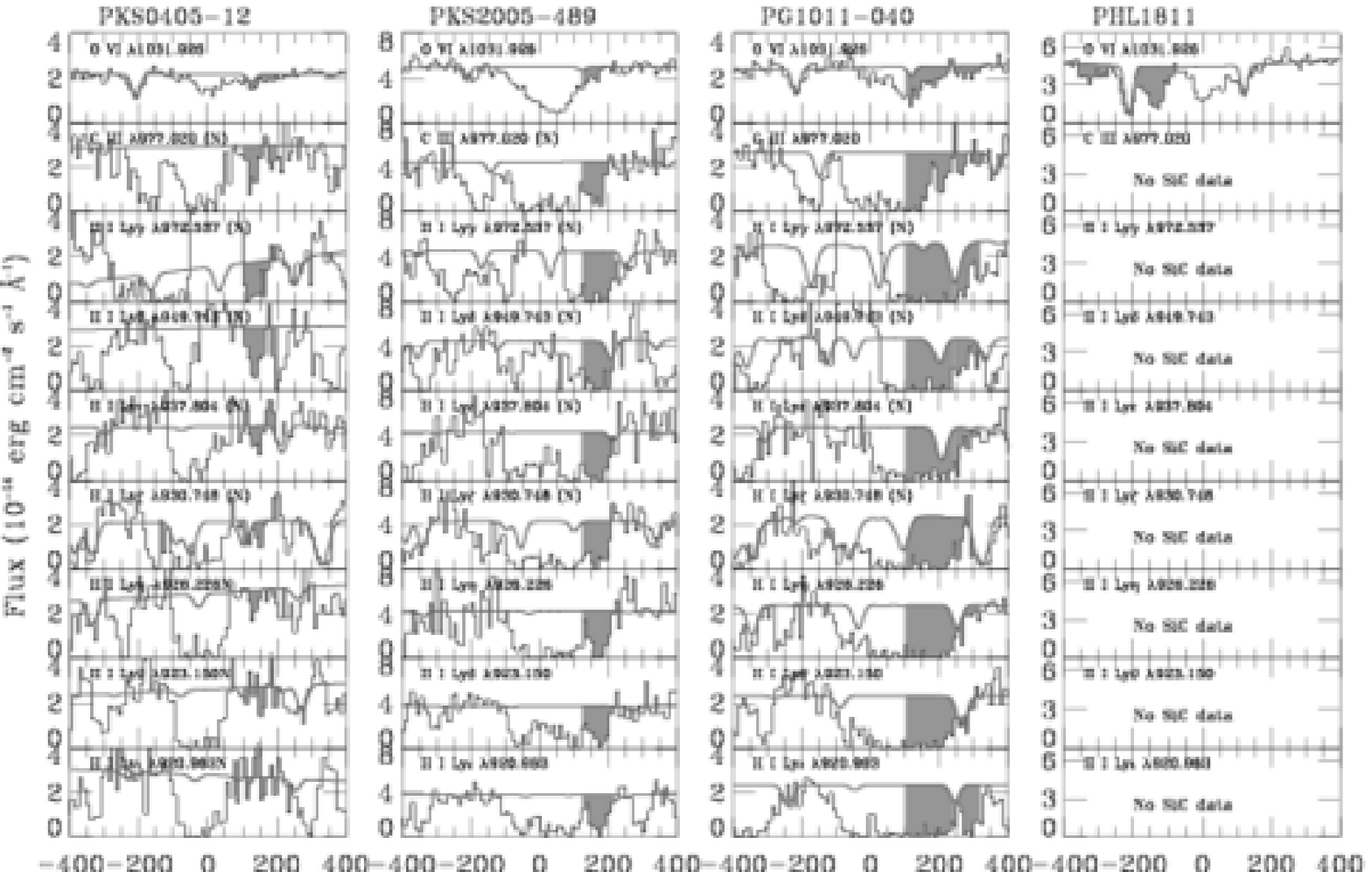}
\plotone{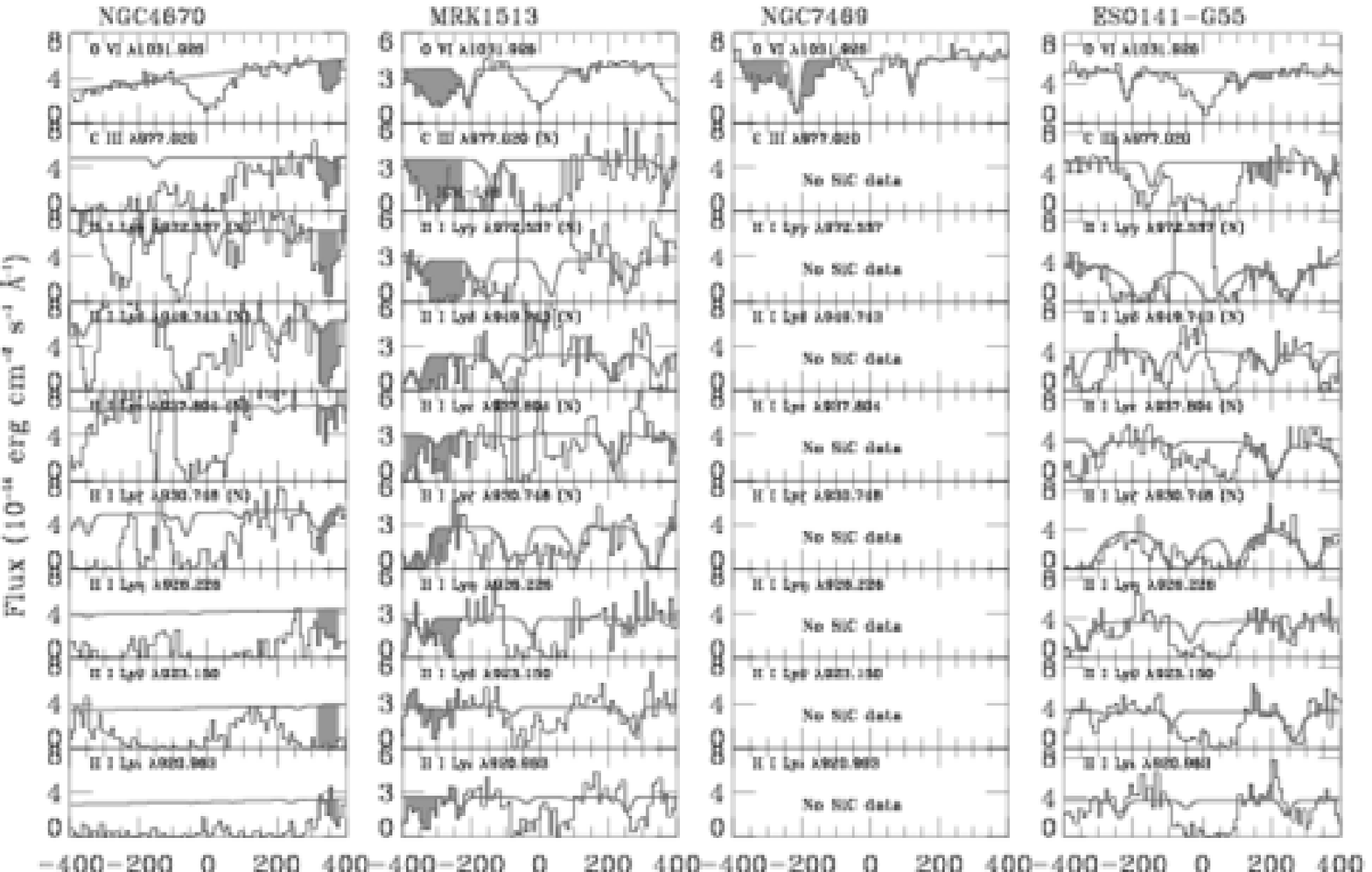}
\caption{}
\end{figure}

\begin{figure}[ht]
\figurenum{2 cont}
\epsscale{1.0}
\plotone{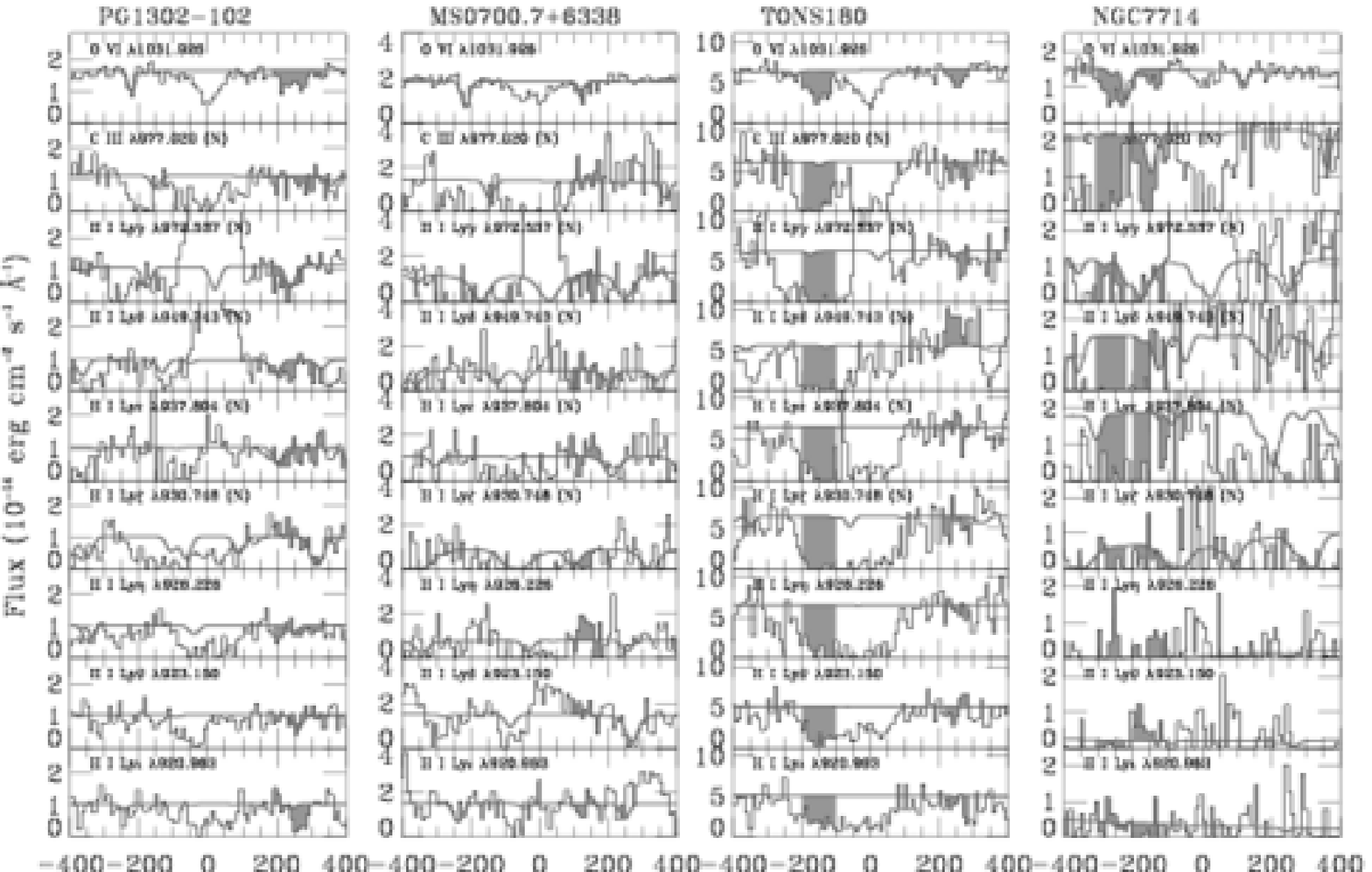}
\plotone{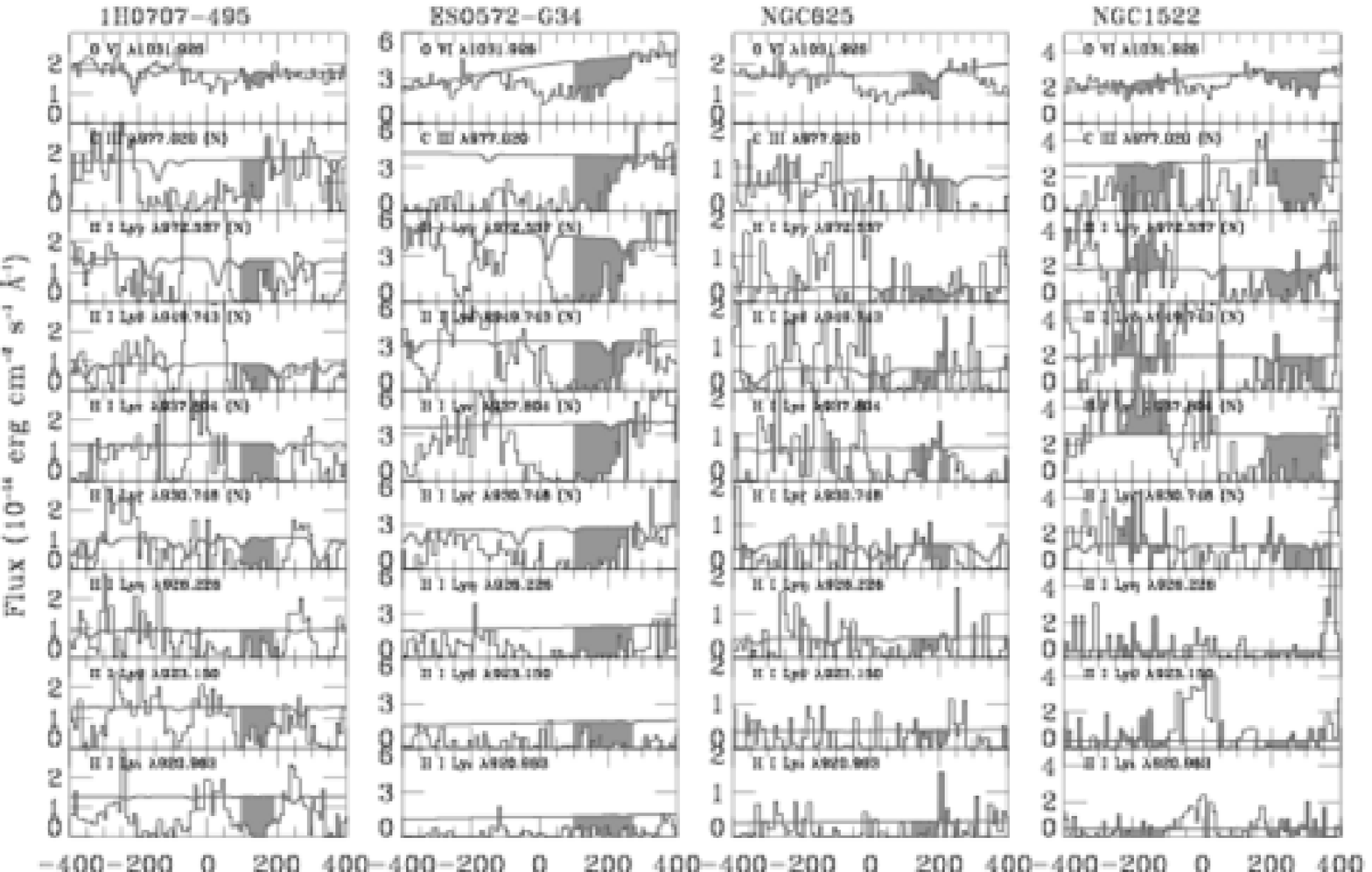}
\caption{}
\end{figure}

\begin{figure}[ht]
\figurenum{2 cont}
\epsscale{1.0}
\plotone{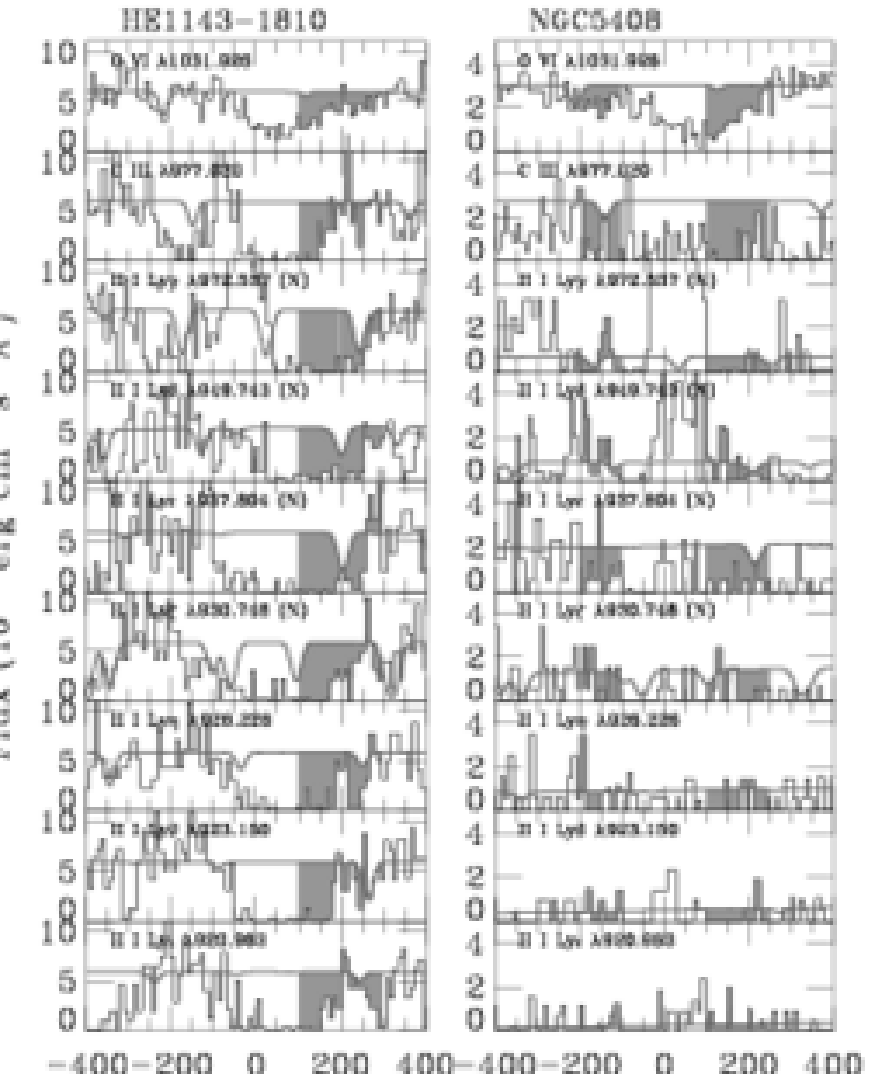}
\caption{}
\end{figure}

\clearpage
\begin{figure}[ht]
\epsscale{0.88}
\plotone{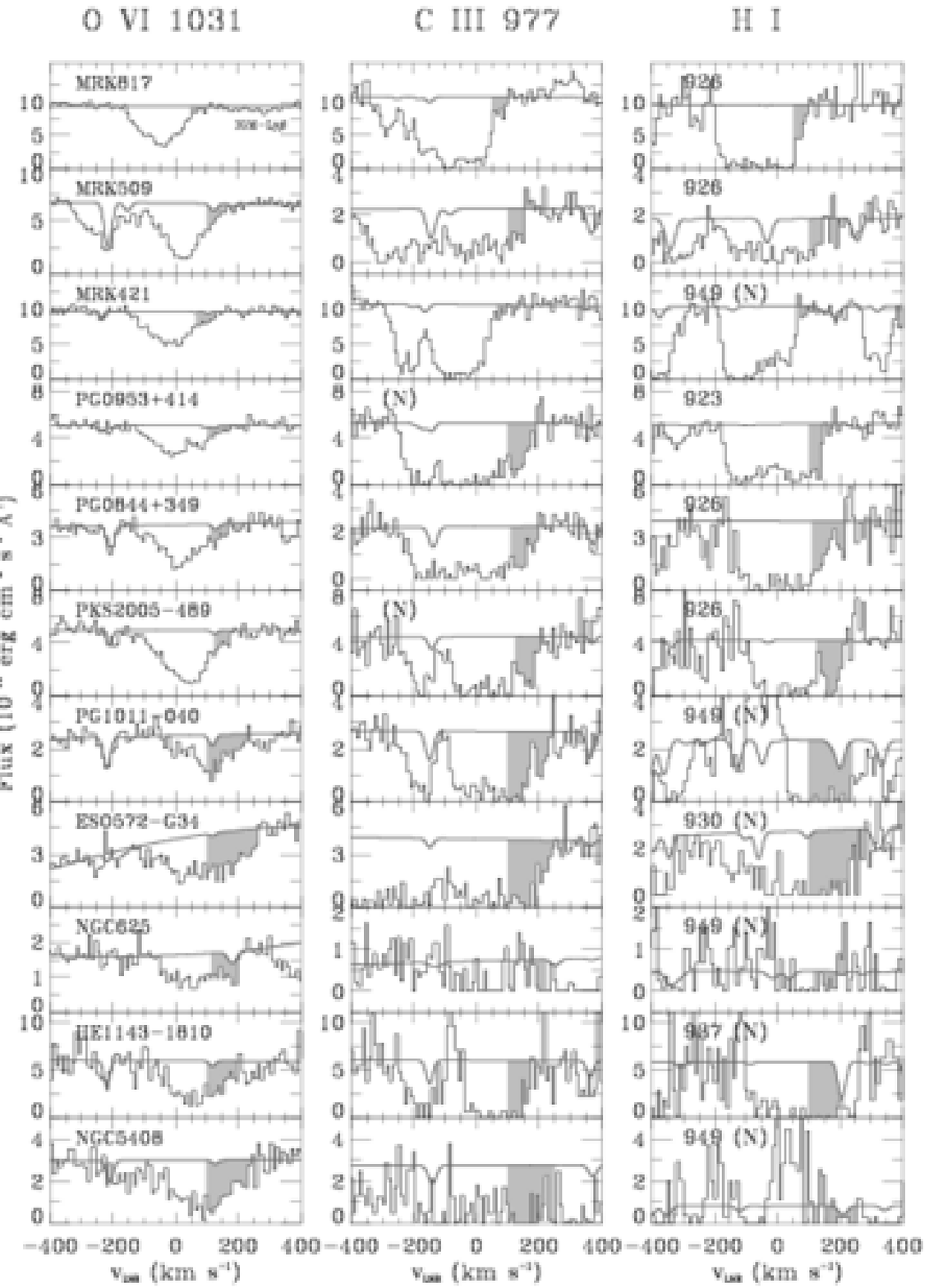}
\caption{\os, \ct, and \hi\ for all \os-detected \pvw s, presented in
  order of decreasing S/N.} 
\end{figure}

\begin{figure}[ht]
\epsscale{0.88}
\plotone{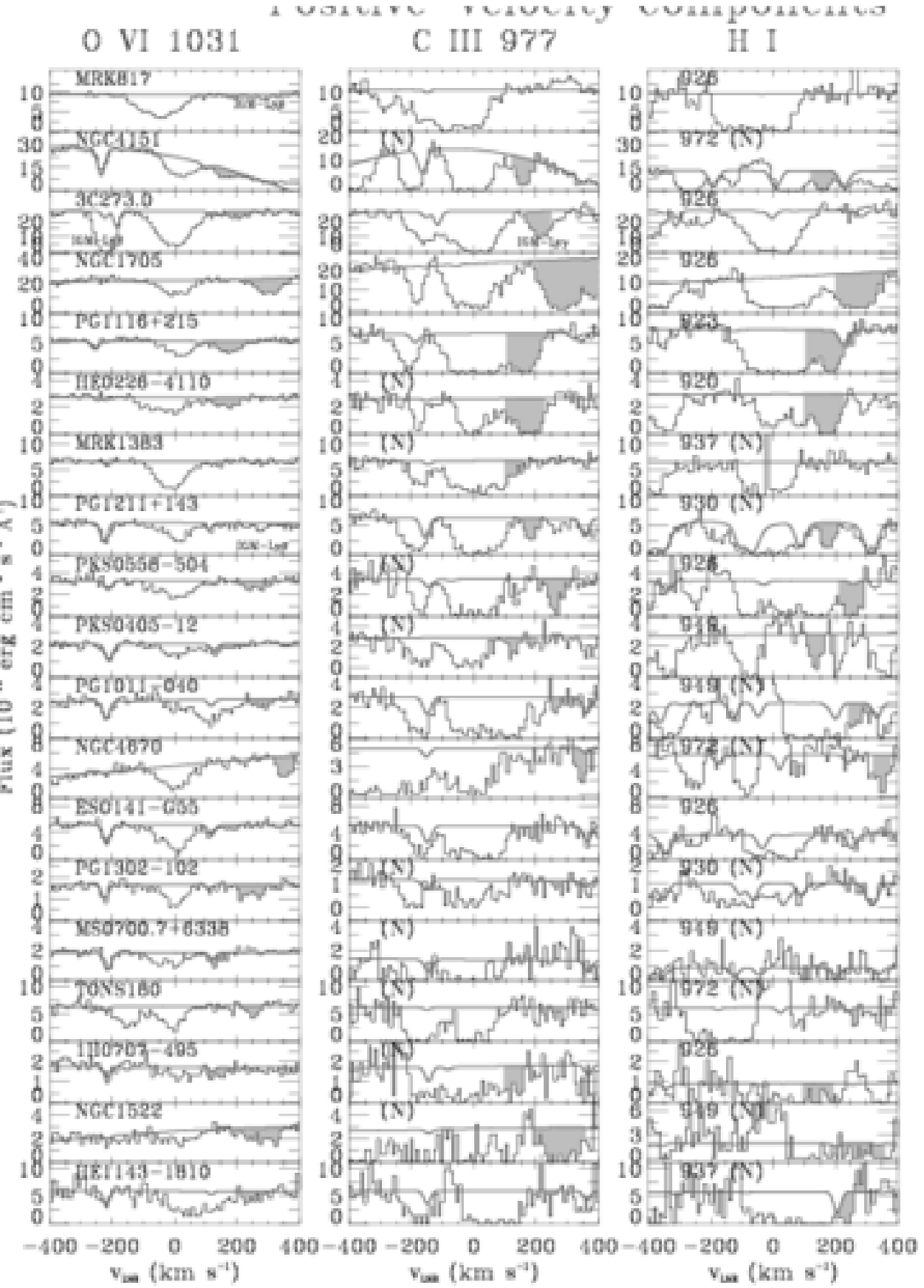}
\caption{Same as Figure 3, but for all \os-detected \pvc s.}
\end{figure}

\begin{figure}[ht]
\epsscale{0.88}
\plotone{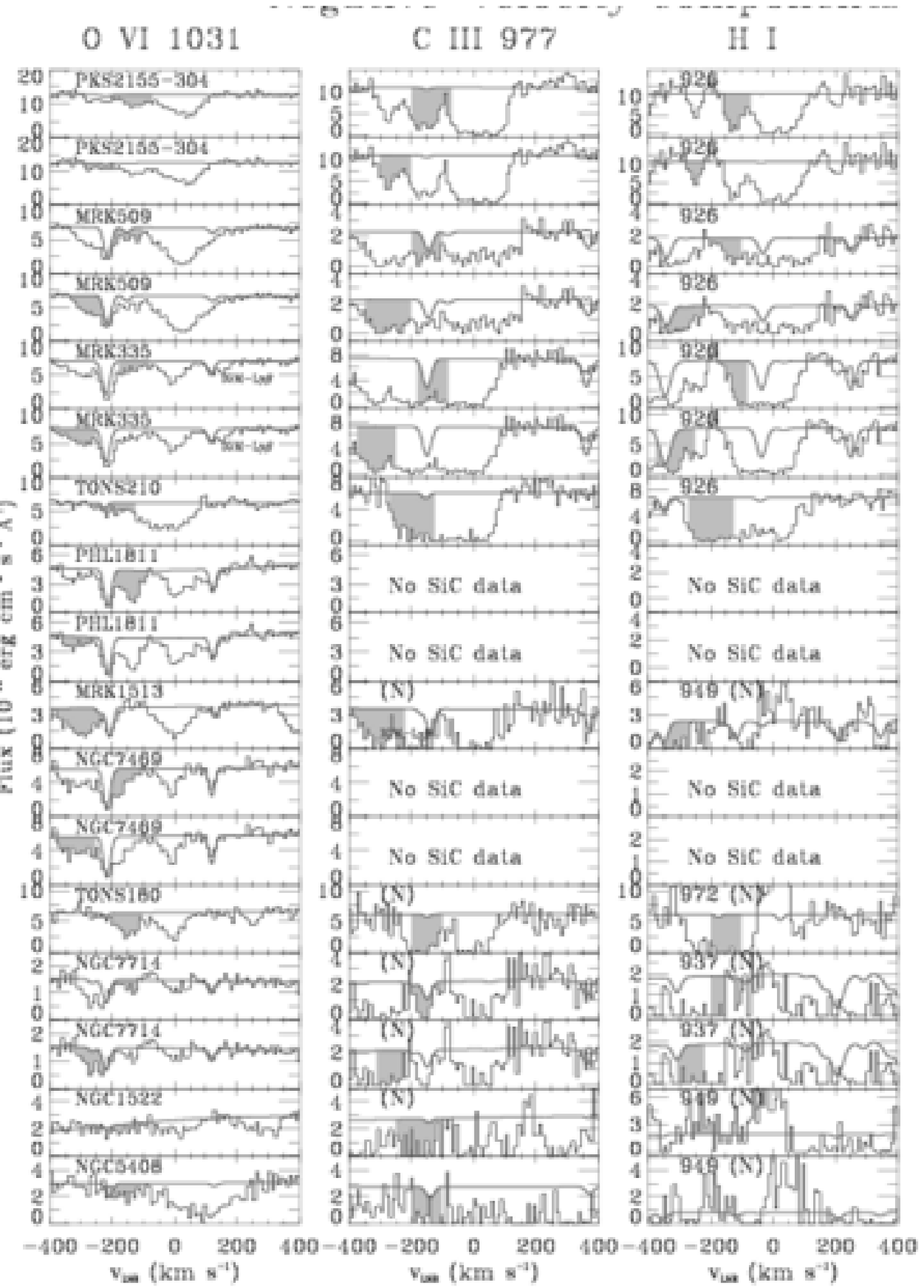}
\caption{Same as Figure 3, but for all \os-detected \nvc s.}
\end{figure}

\begin{figure}[ht]
\epsscale{0.88}
\plotone{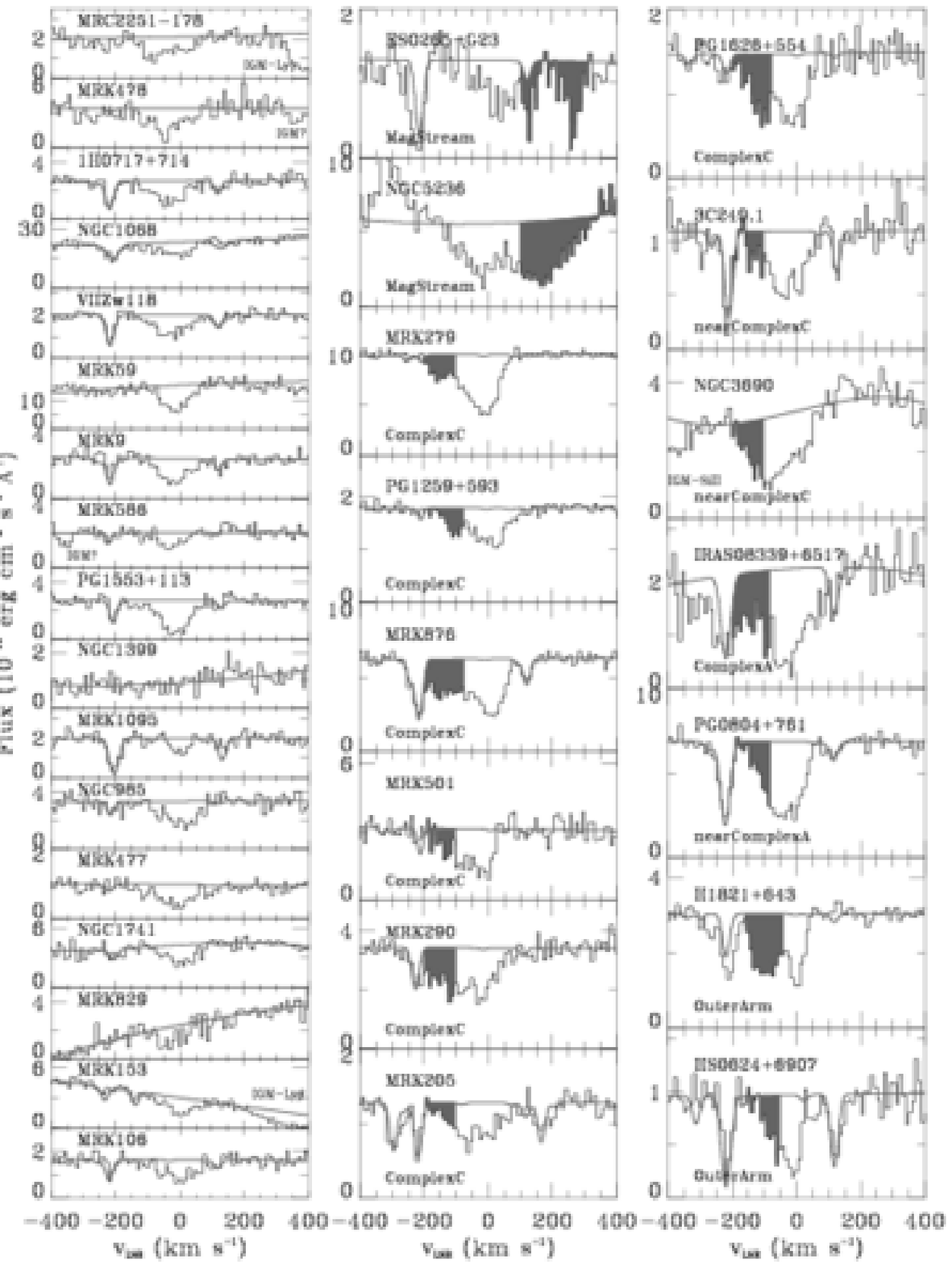}
\caption{\Hv\ \os\ non-detections (left column) and \hvo\ associated with
  21\,cm-emitting HVCs (and hence excluded from our survey; right two
  columns).} 
\end{figure}

\begin{figure}[ht]
\epsscale{0.85}
\plotone{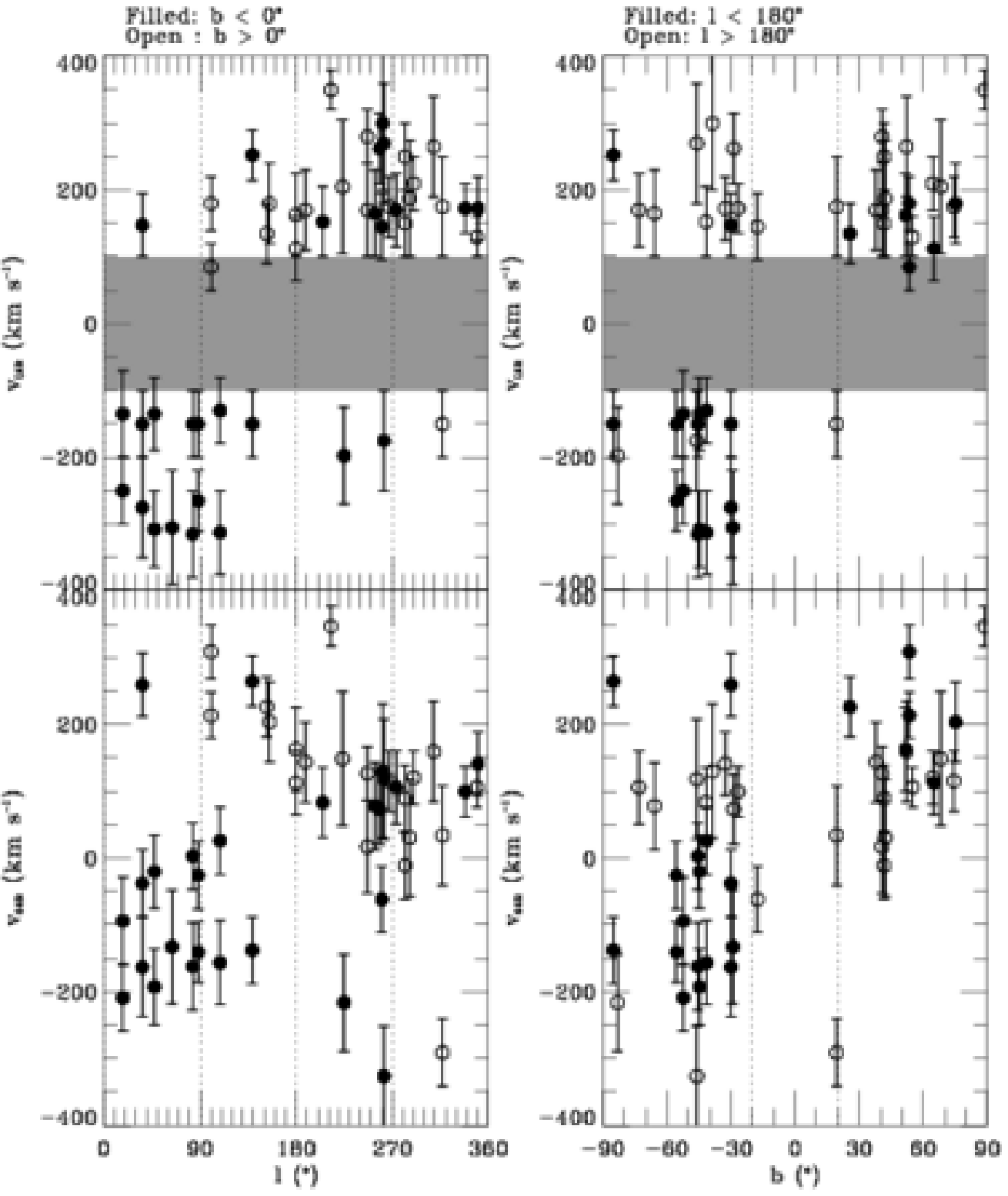}
\caption{Kinematic analysis of \hvo\ absorbers as a
  function of location on the sky. In the top
  panels, $v_{LSR}$ is plotted against Galactic longitude and
  latitude, respectively, with error bars denoting the minimum and
  maximum velocity of absorption. Shading denotes regions with
  $|v_{LSR}|<100$\kms, by definition excluded from our search for 
  high-velocity absorption.
  In the bottom panels, the velocities
  have been converted into the Galactic Standard of Rest reference
  frame.}
\end{figure}

\begin{figure}[ht]
\epsscale{0.85}
\plotone{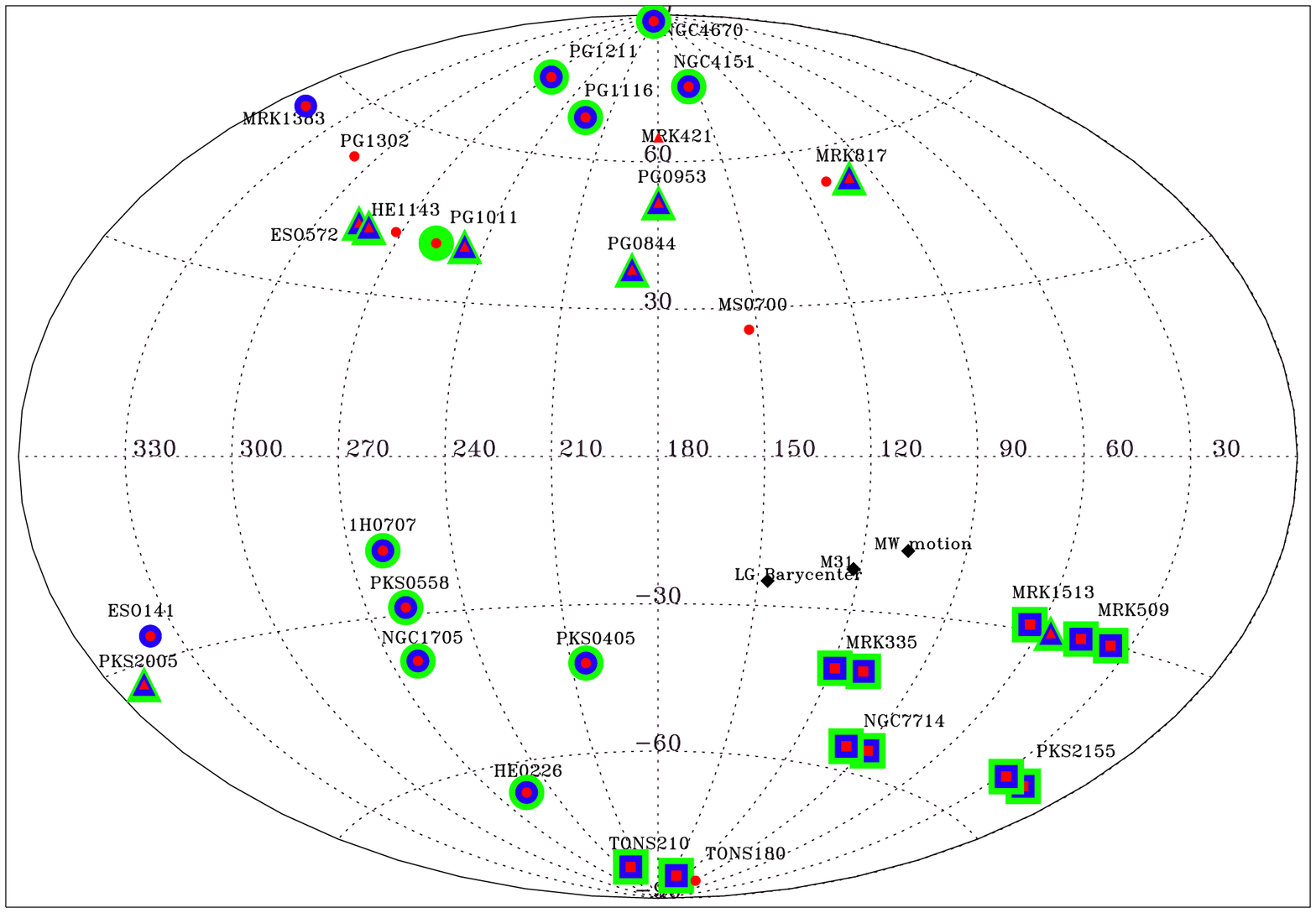}
\caption{Map showing the prevalence of \ct\ and \hi\ absorption
  accompanying \hvo. For a given \hvr\ identified in \os\ (red), a blue
  symbol shows a detection of \ct, and a green
  symbol shows a detection of \hi, in the same velocity range as the
  \os. Circles denote \pvc s, triangles denote \pvw s, and squares denote
  \nvc s. We have omitted cases where blends or poor data quality
  prevent us from knowing whether \ct\ or \hi\ is present.
  In six cases (PKS~2155, Mrk~335, NGC~7714, PG~1011, HE~1143, and
  Ton~S180) there are two sets of symbols denoting two \hv\ absorbers; 
  in one case (Mrk~509) there are three \hv\ absorbers shown. We
  include the following interesting locations: the
  Local Group barycenter \citep[$l, b$=147\degr, $-$25\degr;][]{KM96},
  M~31 ($l,b$=121.17\degr,$-$21.57\degr), and the direction of the Milky
  Way's motion relative to the Local Group barycenter
  \citep[$l,b$=107\degr, $-$18\degr;][]{EL82}.}
\end{figure}

\begin{figure}[ht]
\epsscale{0.8}
\plotone{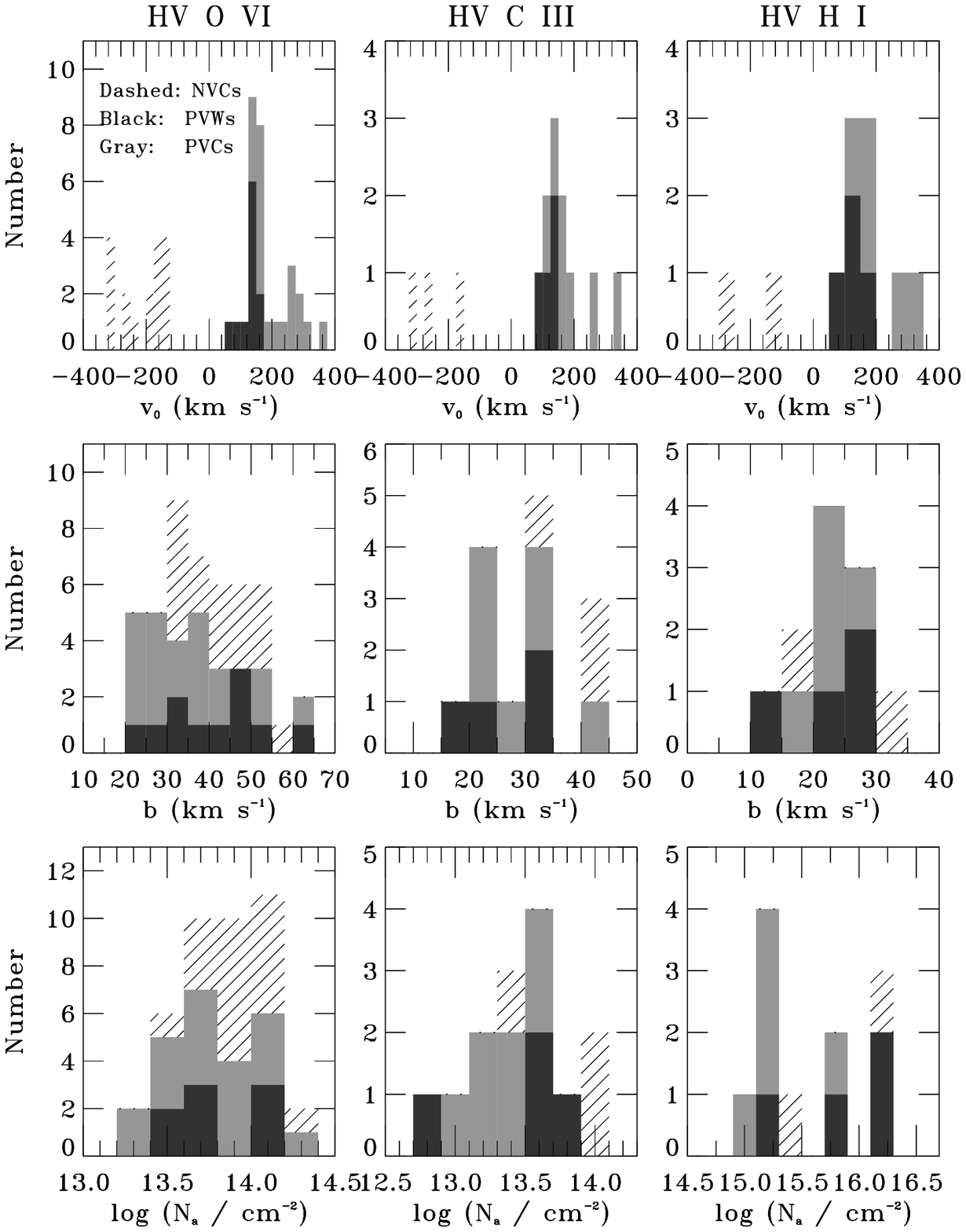}
\caption{Distribution of properties of \os\ HVCs (left column), \ct\ HVCs
  (center column), and \hi\ HVCs (right column). The top row shows the
  distributions of $v_0$, the middle row shows the
  distributions of $b$, and the bottom row shows the
  distributions of log\,$N_a$. \Pvw s are shaded
  in black, \pvc s are in gray, and \nvc s are hatched. Only
  unsaturated \ct\ and \hi\ absorbers are included in this plot.}
\end{figure}

\begin{figure}[ht]
\epsscale{0.8}
\plotone{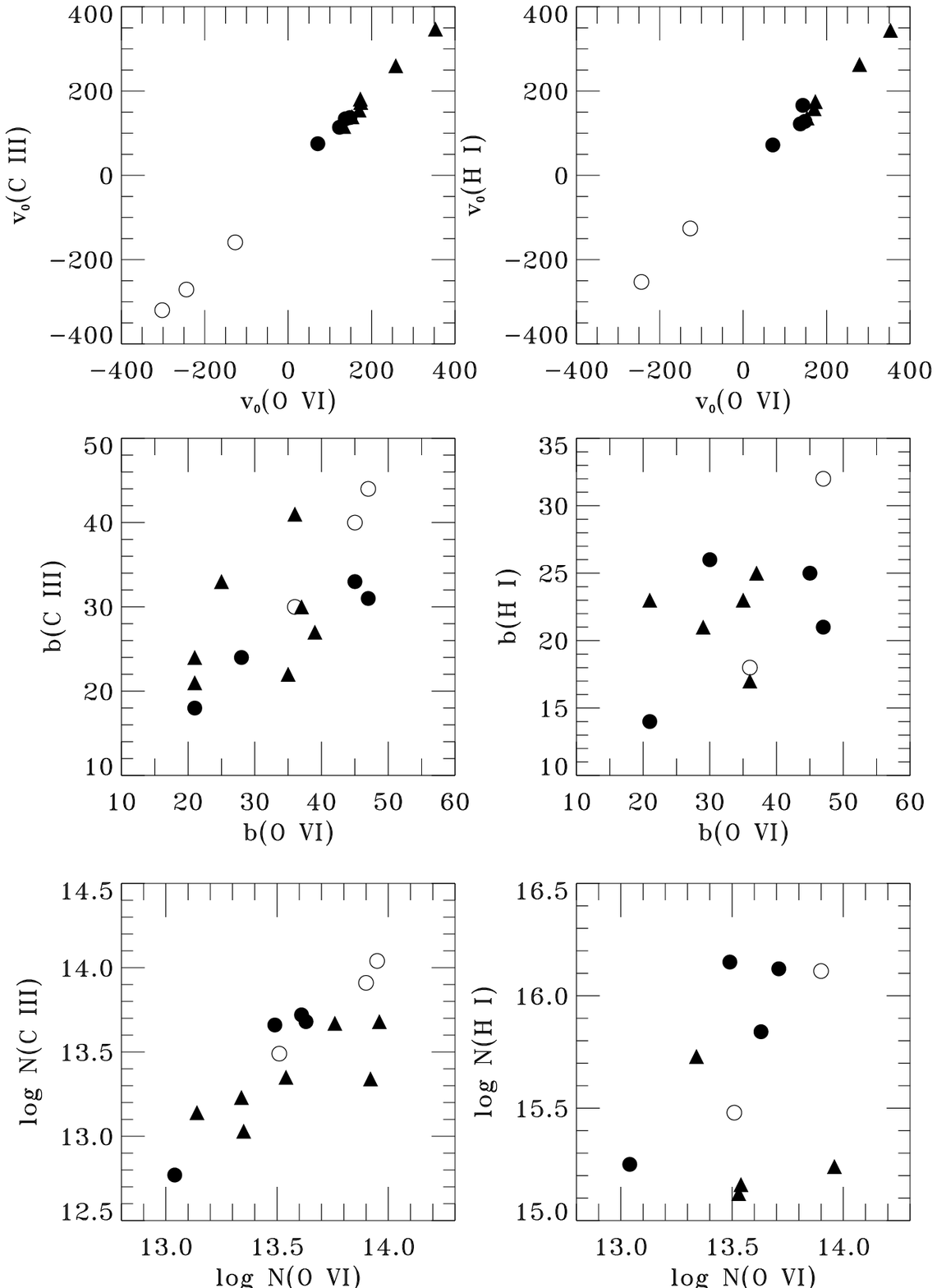}
\caption{Scatter plots comparing properties of the \hvo\ absorbers
  ($v_0$, $b$, log\,$N$) with the corresponding properties in \hv\
  \ct\ absorbers (left panels) and \hvh\ absorbers (right
  panels). Filled circles represent \pvw s, open circles represent
  \nvc s, and filled triangles represent \pvc s. Only absorbers with
  unsaturated \ct\ and \hi\ lines are included in this plot.}  
\end{figure}

\begin{figure}[ht]
\epsscale{0.6}
\plotone{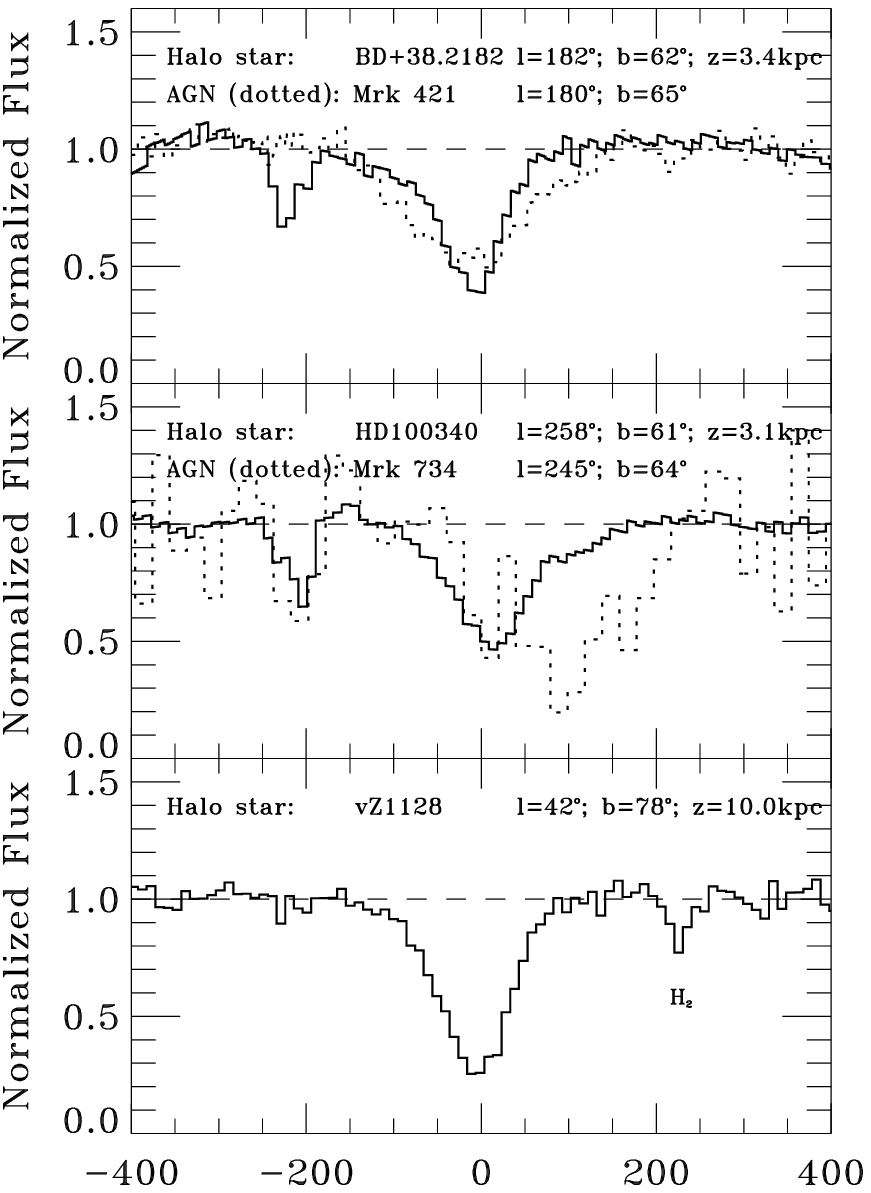}
\caption{\os\ profiles of three high-$z$ halo
  stars and nearby extragalactic sight lines. Dashed lines
  show the continuum position. HD~100340 (solid line,
  middle panel) shows a weak \os\ \pvw\ in the range 80--160\kms,
  whereas BD+38.2182 and vZ~1128 do 
  not. The dotted line in the top panel shows the \os\
  profile along the nearby sight line to AGN Mrk~421 (3.0\degr\
  away from BD+38.2182). The dotted line in the center panel shows the
  \os\ profile 
  toward AGN Mrk~734 (7.0\degr\ away from HD~100340. The presence of
  \hpv\ absorption in the AGN spectra but not in the halo star spectra 
  establishes that the wing absorption in these cases arises beyond
  the halo stars.
  The {\it negative}-velocity wing that appears to be present in the
  BD+38~2182 \os\ spectrum is unrelated to the discussion of
  outflowing gas.}
\end{figure}

\begin{figure}[ht]
\epsscale{0.4}
\plotone{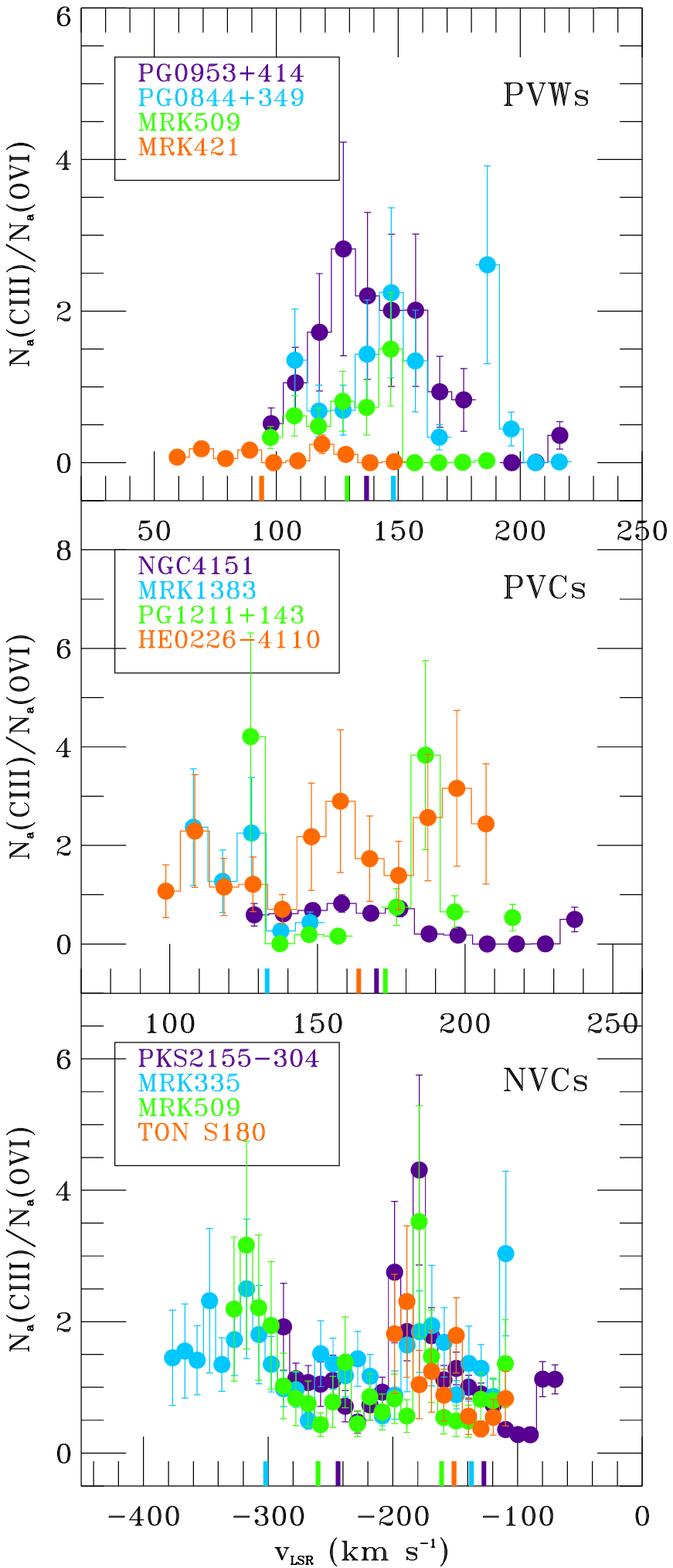}
\caption{Dependence on velocity of the $N$(\ct)/$N$(\os) ratio in \pvw
  s (top), \pvc s (middle), and \nvc s (bottom). Each absorber is
  plotted using a unique color over the velocity range of
  \hva. Absorbers showing little saturation in \ct\ were included in
  this plot. Small vertical marks at the bottom each panel indicate
  the component center of each absorber. With some exceptions, these
  ratios show little slope with $v_{LSR}$.
  If the \ct\ and \os\ exist  
  in the same gaseous phase, the observed ratios of between 0.1 and 4
  imply a CIE temperature of 1.6--1.9$\times10^5$\,K, assuming solar
  abundances.}
\end{figure}


\begin{figure}[ht]
\epsscale{0.80}
\plotone{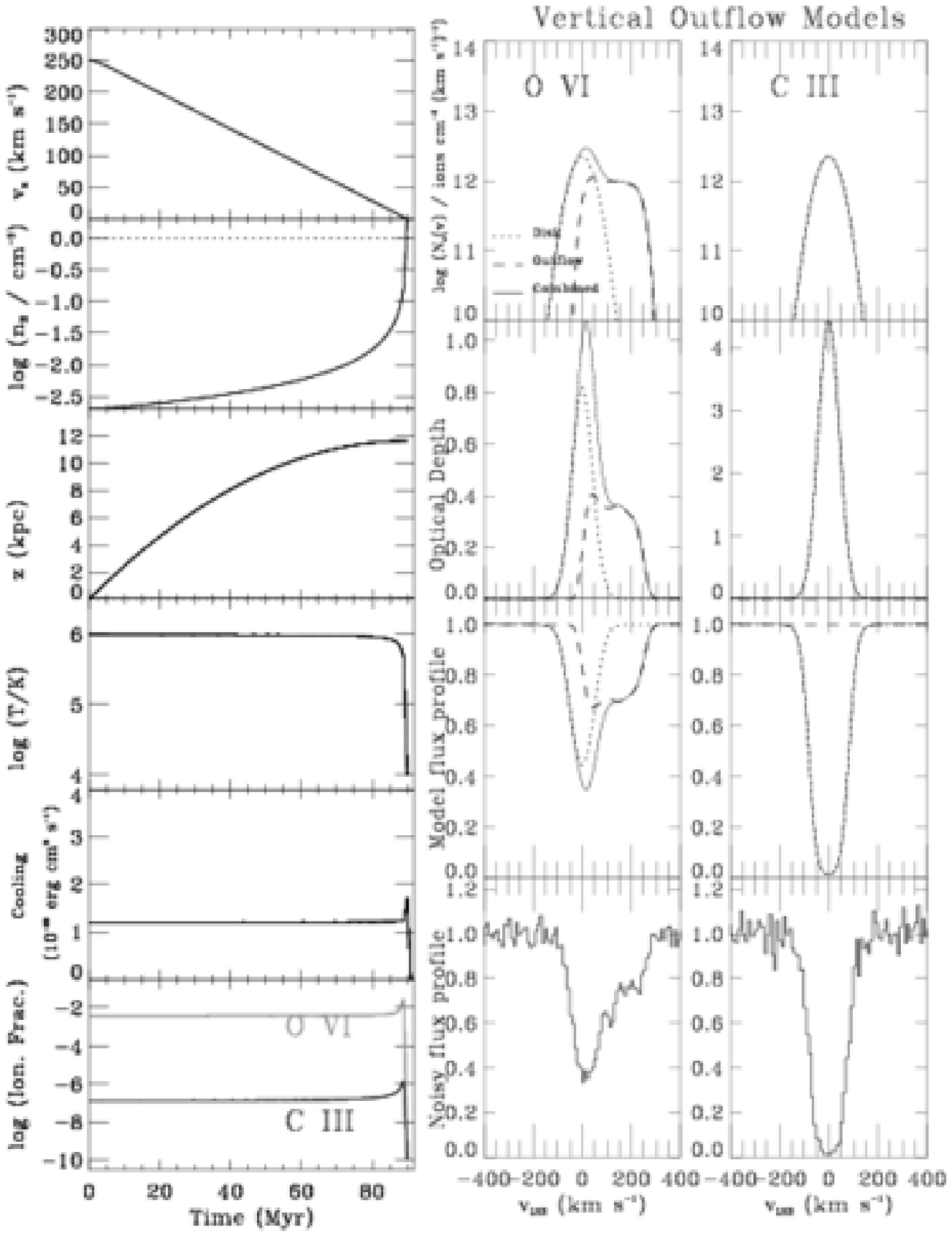}
\caption{Results of modelling a radiatively cooling vertical Galactic
  outflow with dynamics dominated by the Galactic
  gravitational field, for the case $b$=90\degr. 
  This model run has $n_0=2.0\times10^{\-3}$\,cm$^{-3}$, $v_0=250$\kms, and
  $T_0=10^6$\,K. The left panels show the run of various physical
  properties of the flow with time in Myrs. 
  The right panels show the predicted \os\ and \ct\ 
  profiles, in column density per unit velocity, optical depth, flux after
  instrumental broadening, and flux with Poisson Noise
  introduced at a level of S/N = 15. The dashed
  line shows the outflow component, the dotted line shows a thick-disk
  component \citep{Sa03}, and the solid line shows the sum of the two. The wing on
  the \os\ profile traces the higher-velocity, higher-temperature stages
  of the outflow. Little \ct\ exists at these temperatures in CIE, so
  our outflow modelling does not predict a wing on the \ct\ profile.}
\end{figure}



\begin{figure}[ht]
\epsscale{0.85}
\plotone{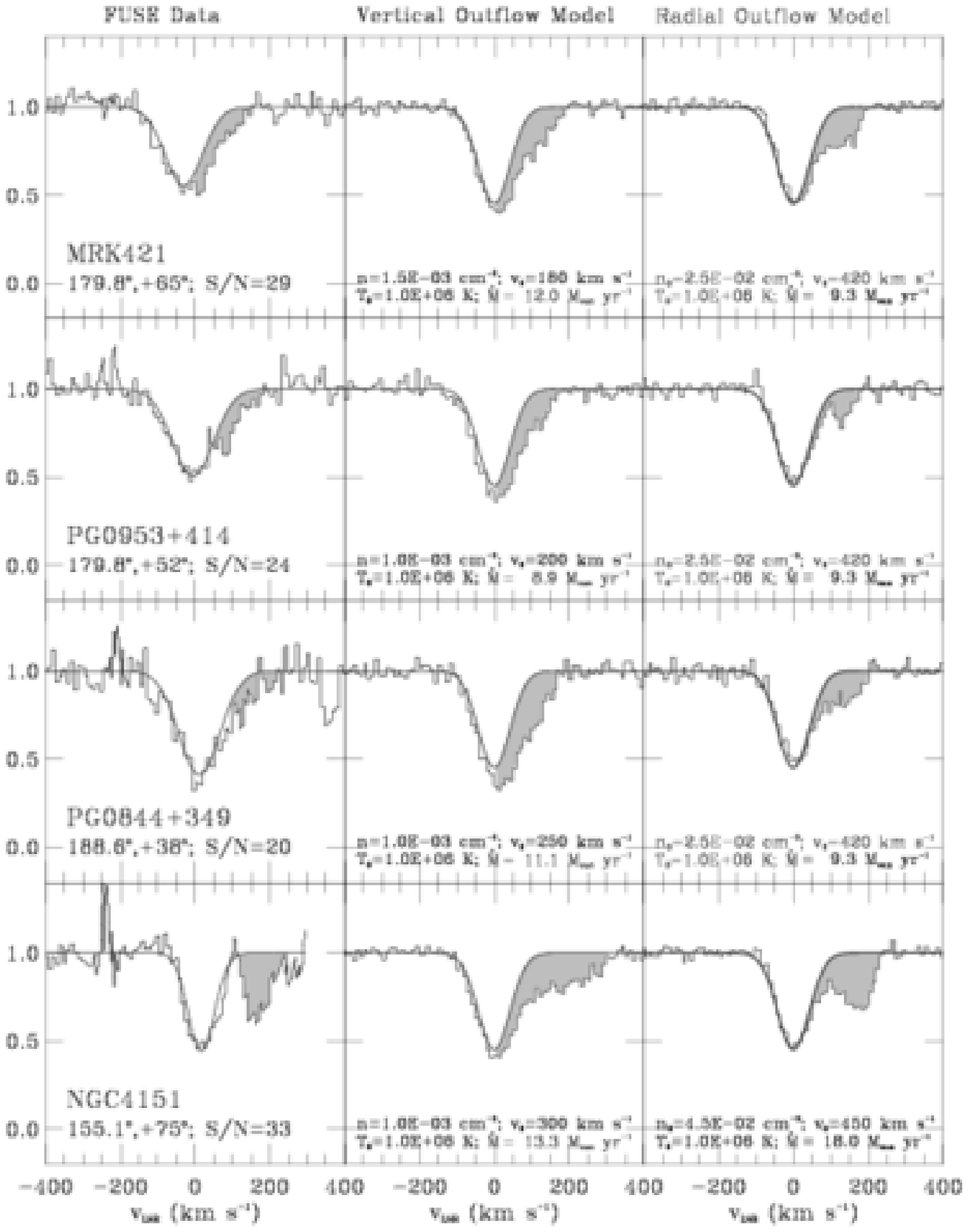}
\caption{Comparison of four observed \hvo\ absorbers (left) near
  $l$=180\degr\
  with the simulated results from vertical (center) and radial (right)
  outflow models. Noise has been added to the model profiles at the
  same level as in the data. In each panel, the smooth line shows the
  thick disk component,  and the shading
  shows the residual, \hv\ absorption. The vertical outflow models
  produce broad \pvw s, whereas the radial
  outflows produce narrower \os\ absorption components, due to
  ``velocity-crowding'' projection effects.}
\end{figure}

\end{document}